\documentclass[prl,aps,twocolumn,showpacs,floatfix]{revtex4-1}

\usepackage{dcolumn}
\usepackage{amsmath}
\usepackage{amssymb}
\usepackage{bm}
\usepackage{graphicx}
\usepackage{epsf,epsfig}
\usepackage{hyperref}
\usepackage{color}

\begin{document}
\title{How does the extracellular matrix affect the rigidity of an embedded spheroid?}
\author{Amanda Parker$^1$, M. Cristina Marchetti$^2$, M. Lisa Manning$^1$, J. M. Schwarz$^{1,3}$}
\affiliation{$^1$Department of Physics and BioInspired Syracuse, Syracuse University, Syracuse, NY 13244, USA
$^2$Department of Physics, University of California at Santa Barbara, Santa Barbara, CA, 93106, USA
$^3$Indian Creek Farm, Ithaca, NY, 14850, USA}
\date{\today}

\begin{abstract}

Cellularized tissue and polymer networks can both transition from floppy to rigid as a function of their control parameters, and, yet, the two systems often mechanically interact, which may affect their respective rigidities. To study this interaction, we consider a vertex model with interfacial tension (a spheroid) embedded in a spring network in two dimensions. We identify two regimes with different global spheroid shapes, governed by the pressure resulting from competition between interfacial tension and tension in the network. In the first regime, the tissue remains compact, while in the second, a cavitation-like instability leads to the emergence of gaps at the tissue-network interface. Intriguingly, compression of the tissue promotes fluidization, while tension promotes cellular alignment and rigidification, with the mechanisms driving rigidification differing on either side of the instability.

\end{abstract}

\maketitle

\noindent \textit{\textbf{Introduction.}}
Cellularized tissue-- groups of cells exhibiting collective behavior-- is held together by adhesive cell-cell junctions, which regulate an incredible range of functions, including wound healing and embryogenesis \cite{Cooper_2000, Niessen_2011, Li_2013, Garcia_2018, Bruckner_2018}. In addition, there is typically a non-cellular component present, the extracellular matrix (ECM)-- a network often largely composed of cross-linked collagen fibers-- to which cells mechanically couple via cell-ECM adhesions. Through these adhesions, the ECM imposes forces on cells to provide structural support and regulate cell-cell interactions, driving changes in cell behavior and, ultimately, tissue mechanics \cite{Galbraith_1998, Lo_2000, Lui_2017, Humphrey_2014, Frantz_2010}. Maintenance of these interactions is crucial for healthy tissue, and, in fact, disruptions in forces exerted by the ECM, or in the cells' ability to sense or respond to mechanical signals from the ECM, are hallmarks of disease \cite{Walker_2018, GIUSSANI_2015, Acerbi_2015, Leight_2017, McKenzie_2018}.

A key \textit{in vitro} model system for studying the combined effects of cell-cell and cell-ECM interactions on tissue behavior is a tissue spheroid embedded in a collagen network. It has been shown that tensile forces and stiffer collagen fibers in the ECM facilitate tumor invasion ~\cite{Kopanska_2016, Suh_2019}, and that as various ECM proteins are modified, cell shapes may be a good predictor of invasion potential~\cite{Baskaran_2020}. 

Despite the recent spate of experimental work on multicellular systems embedded in biopolymer networks, numerical models for cells interacting with ECM have focused on single cells, potentially missing important features associated with collective cell behavior \cite{Zheng_2019, Kim_2018, Han_2018}. Yet, models for bulk tissue and collagen network have separately been quite successful. One such model is a cell-based vertex model \cite{Honda_2001, Farhadifar_2007,Bi_2015}, which predicts a density-independent rigidity transition in disordered confluent tissues and micro-demixing in tissue mixtures ~\cite{Bi_2015,Merkel_2018, Sahu_2019}. Both phenomena have been verified experimentally \cite{Park_2015, Malinverno_2017, Sahu_2019}. Meanwhile, the strain-stiffening behavior observed in biopolymer networks has been captured by under-constrained spring network models \cite{Licup_2015, Jansen_2018}. Interestingly, in the absence of energetic penalties for bending, the spring network's floppy-to-rigid transition at finite shear strain is yet another density-independent rigidity transition in the same universality class as that of the vertex model \cite{Merkel_2019}.

Here we couple a tissue-based vertex model to a surrounding spring network in two dimensions with line tension at the interface, and study the rheology and morphology of the tissue. Our model is minimal as our goal is to investigate the effect of mechanical coupling alone on the collective behavior of pre-migratory tissue cells. However, even for this minimal bi-material, without feedback to dynamically up- or down-regulate model parameters, nontrivial changes in cell shape, tissue phase, and overall tissue structure can be induced. Using mean-field approximations, we analytically quantify the state of the tissue in two distinct regimes, separated by an instability in the tissue boundary, to predict the fluidity and geometry of the tissue cells, using arguments that are generalizable to any biological system where the competition between external forces and interfacial tension plays an important role.

\begin{figure}
    \centering
    \includegraphics[width=0.43\textwidth]{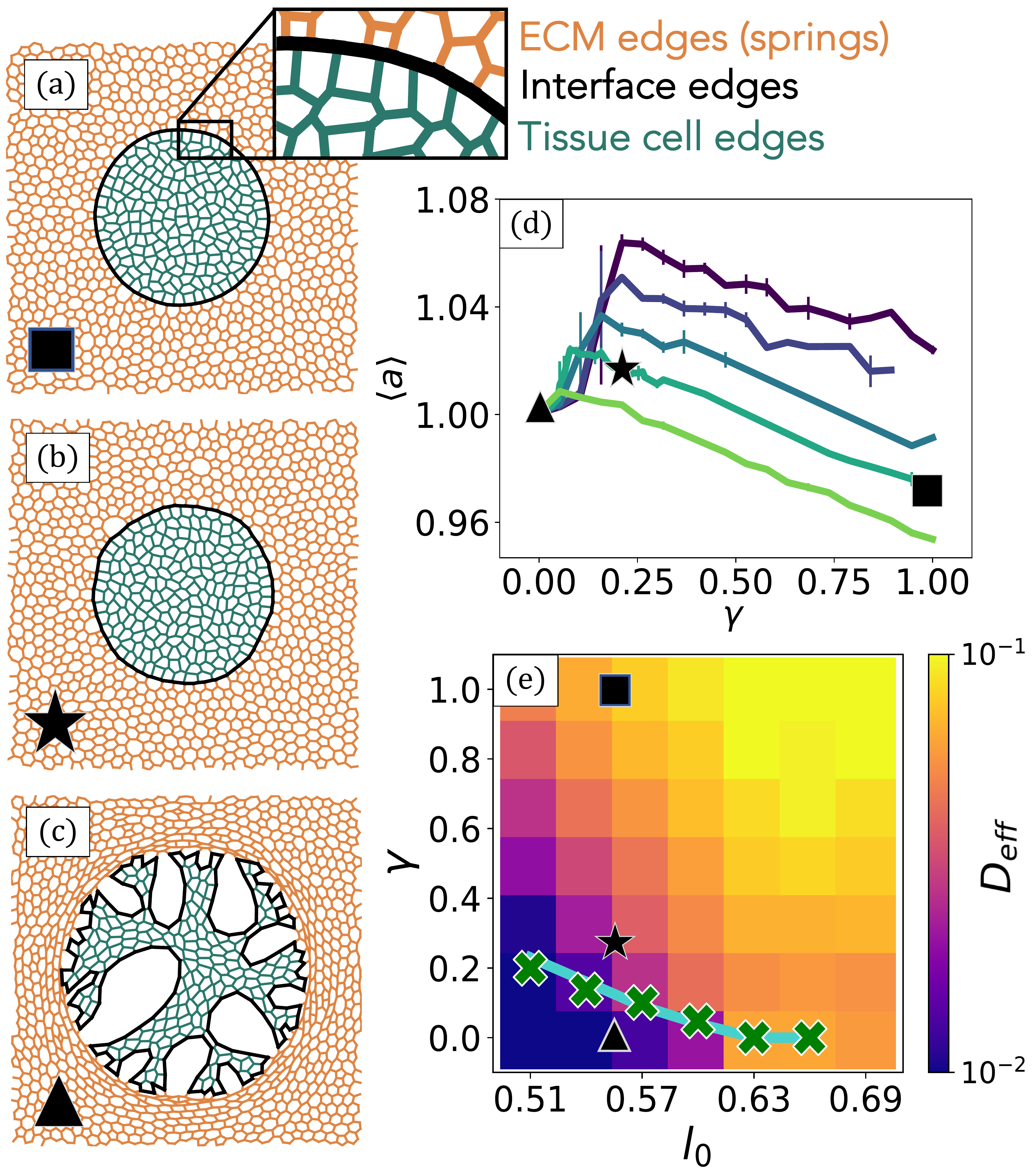}
    \caption{\textit{Tissue configurations, average cell shape, and fluidity, for fixed $p_{0}=3.91$.}
    (a)-(c) Minimal energy configurations for fixed equilibrium spring length, $l_0=0.56$, and varying interfacial tension, $\gamma = $ 1.0, 0.22, and 0.0, including a detail showing springs in orange, interface edges in black, and cell edges in green. (d) The average tissue cell area, $\langle a \rangle$, versus $\gamma$ for a series of $l_{0}$ values equally-spaced from $0.45$ (purple) to $0.6$ (light green). (e) A heatmap of the effective diffusion constant, $D_{eff}$, of tissue cells as a function of $\gamma$ and $l_0$. The line indicates the analytical prediction for the cavitation instability, $\gamma_{c}^{MF}$, while Xs denote the onset of the instability, $\gamma_{c}$, measured from simulations. The black symbols correspond to the same three sets of parameters throughout.}
    \label{fig_full_model}
\end{figure}

\noindent {\it \textbf{Model.}} 
Our model is composed of a tissue of $N_{cells}$ cells, each with area, $a_\alpha$, and perimeter, $p_\alpha$, embedded in a network of springs, each with equilibrium spring length, $l_0$, and stiffness, $k_{sp}$, with line tension, $\gamma$, at the interface (see Fig. \ref{fig_full_model} (a), detail). The dimensionless energy functional is 
 \begin{multline}
    e_{total} = \sum_{cells, \alpha} (a_{\alpha} - 1)^{2} + k_{p} \sum_{cells, \alpha} (p_{\alpha} - p_{0})^{2} \\ 
     + \frac{1}{2} k_{sp} \sum_{sp\langle ij \rangle}  (l_{sp \langle ij \rangle} - l_{0})^{2} + \gamma \sum_{int \langle ij \rangle}  l_{int\langle ij \rangle},
 \label{eq_nondim_energy_total_final}
 \end{multline}
with the first two terms representing the standard vertex model for the tissue, the third term the elasticity of the spring network with edge lengths $l_{sp\langle ij\rangle}$, and the fourth term the interfacial tension, with edge lengths $l_{int\langle ij \rangle}$.  The control parameter for rigidity in the standard vertex model is the preferred cell shape index, $p_{0}$, with values of $p_{0}$ below a threshold inducing rigidity in the cellularized tissue (see Supplementary Material (SM) Sec. A). In a homogeneous, under-constrained spring network system, the control parameter for rigidity (tension) is the rest length of the springs, $l_0$, with a rest length below some threshold resulting in a rigid network (see SM Sec. B). For simplicity, we study the spring network in the limit of zero bending energy where the rigid-to-floppy transition is well-defined. 

Using zero-temperature energy-minimization and low-temperature Brownian dynamics (see SM Sec. D), we explore minimal-energy configurations of this model, focusing on the emergent morphology and rheology of the tissue. In doing so, the rules for tissue cell T1 transitions near the boundary must be carefully considered (see SM Sec. C).

{}

\noindent {\it \textbf{Results.}} We find that both the phase and morphology of the tissue depend on the balance of interfacial tension, $\gamma$, and the equilibrium spring length, $l_{0}$, of the surrounding network, for a fixed $p_{0}$ of the tissue cells. As $l_0$ decreases, the mean spring length, $\langle l_{sp} \rangle$, decreases to minimize the total energy. This decrease necessitates the shrinking of the area taken up by the springs, either by increasing the size of the tissue or by opening gaps at the tissue-ECM interface. At the same time, changes to the global area of the tissue spheroid cost energy due to a quadratic penalty for deviations from the preferred cell area, while changes to the tissue-ECM interface are penalized by the interfacial tension and induced deviations from the preferred cell perimeter. 

The competition between these effects is illustrated in Fig. \ref{fig_full_model} for $p_{0}=3.91$. The configurations in (a)-(c) depict the effect on the overall tissue shape of decreasing $\gamma$ ($\gamma_{(a)}=1.0$, $\gamma_{(b)}=0.22$, $\gamma_{(c)}=0.0$) for fixed $l_{0}$. We observe roughly circular shapes for the tissue in the first two cases, while at $\gamma=0$ cavities appear, yielding a different global shape. Fig. \ref{fig_full_model} (d) shows the mean cell area, $\langle a \rangle$, as a function of $\gamma$, for varying $l_{0}$ values (see caption). Fig. \ref{fig_full_model} (e) shows the effects of $\gamma$ and $l_{0}$ on the tissue fluidity, quantified by the effective diffusion coefficient, $D_{eff}$ (see SM Sec. D2), where higher values correspond to more fluid-like tissues. Points with the same parameters as in (a)-(c) are marked with corresponding symbols in (d) and (e). 

For a fixed $l_{0}$ corresponding to rigid ECM, increasing $\gamma$ results in a smooth increase in $D_{eff}$, as shown in (e), while $\langle a \rangle$ changes non-monotonically, as shown in (d). As we increase $\gamma$ beyond some critical value, $\gamma_c$, defined by the maximum of each curve in (d), $\langle a \rangle$ decreases, while the overall tissue geometry remains roughly circular. As we decrease $\gamma$ below $\gamma_c$, $\langle a \rangle$ also decreases, but ECM tension induces cavities. This behavior is indicative of an instability, whose location we predict analytically and plot as a solid line in (e), while the Xs denote the locations of $\gamma_c$ determined numerically.

\noindent {\it \textbf{Cavitation Instability.}} To analytically predict the onset of the instability occurring at $\gamma_c$, we devise a mean-field model in which a circular tissue containing a circular cavity is embedded in a regular, hexagonal spring network. We assume that all cells remain at their preferred area and perimeter, leaving contributions only from the tissue boundary and spring network. This assumption allows us to write the total energy as a function of the cavity radius, $r$, only (see SM Sec. F). For a fixed $l_{0}$ value, we find two energy minima--- one at $r=0$, and another at $r>0$ at a particular value of $\gamma$ which we denote $\gamma_{c}^{MF}$. For $\gamma > \gamma_{c}^{MF}$, the energetically-favorable cavity radius is zero, and the system prefers circular, compact tissue geometries. For $\gamma < \gamma_{c}^{MF}$, the system prefers a cavity of finite radius, which grows as $\gamma$ decreases. The computed $\gamma_{c}^{MF}$ agrees well with the measured $\gamma_{c}$ identified by the location of the peak in the $\langle a \rangle$ vs. $\gamma$ curve for the corresponding $l_{0}$ (see Fig. \ref{fig_full_model} (e) and SM Sec. F). As $l_0$ decreases, $\gamma_{c}^{MF}$ increases, since larger interfacial tension is required to overcome the increased tension in the network and prevent the formation of cavities.

\begin{figure}
    \centering
    \includegraphics[width=0.41\textwidth]{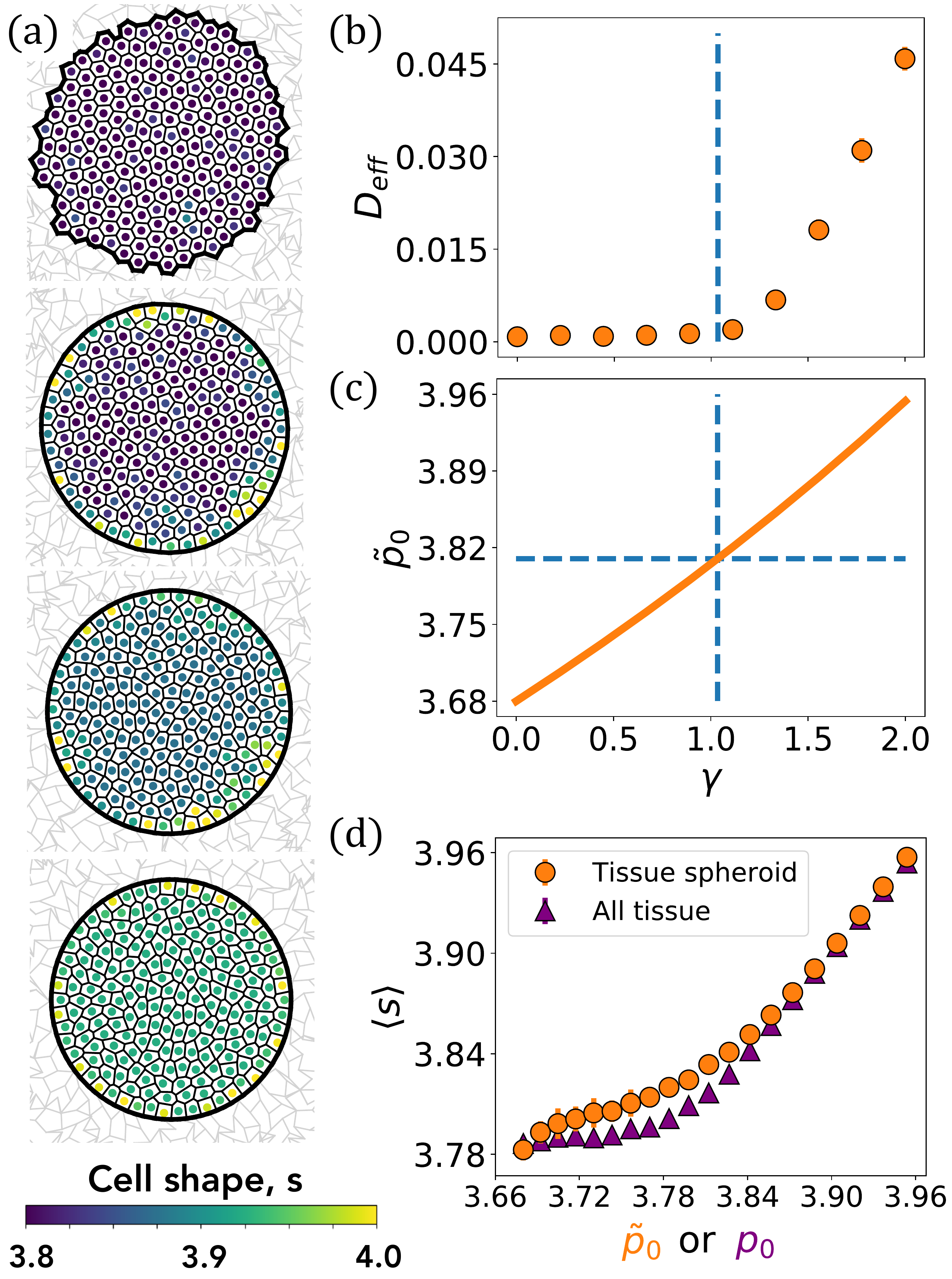}
    \caption{\textit{Compact tissue regime results for fixed $p_{0}=3.68$.} (a) Minimum-energy configurations for a series of interfacial tension, $\gamma$, values. From top to bottom, $\gamma$=0.0, 0.53, 1.47, and 1.79. Each cell center is colored by the cell's shape, $s = p/\sqrt{a}$, where $p$ and $a$ are the cell's perimeter and area. (b) The effective diffusion coefficient, $D_{eff}$, vs. $\gamma$. Below a critical value of $\gamma$, $D_{eff} \approx 0$ and the tissue is solid-like, and above, $D_{eff} > 0$ and the tissue is fluid-like. The $\gamma$ value at which the effective target shape index, $\tilde{p}_{0}$, is equal to 3.81 is indicated by a vertical dashed line. (c) The effective target shape index, $\tilde{p}_{0}$, as a function of $\gamma$. The horizontal dashed line denotes $\tilde{p}_{0}=3.81$, and the corresponding $\gamma$ value is denoted with a vertical dashed line. (d) The observed mean cell shape, $\langle s \rangle$, vs. $\tilde{p}_{0}$. Simulation data for varying $\gamma$ is plotted with circles, while data for a bulk vertex model with varying $p_{0}$ is plotted with triangles for comparison.}
    \label{fig_case_01_panel}
\end{figure}

\noindent {\it \textbf{Compact tissue regime.}} When $\gamma > \gamma_{c}$, the tissue is roughly circular, or ``compact,'' with no cavities. To understand this regime, we first focus on the limit of zero ECM tension. Thus, increasing the interfacial tension, $\gamma$, drives the tissue boundary to shrink, reducing the tissue radius and compressing the tissue, as shown in Fig. \ref{fig_case_01_panel} (a). To study the effect of this compression on tissue fluidity, we compute the cells' effective diffusion coefficient, $D_{eff}$, as a function of interfacial tension, $\gamma$. Fig. \ref{fig_case_01_panel} (b) shows results for a tissue with $p_{0}=3.68$. At $\gamma=0$, $D_{eff} \approx 0$, and the tissue is solid-like. As $\gamma$ increases and the tissue cells are compressed, $D_{eff}$ becomes non-zero at a particular value of $\gamma$.

This compression-induced fluidization can be understood via a mapping of the embedded tissue to an effective bulk vertex model. By approximating the tissue geometry as circular, ignoring fluctuations in the cell perimeters, cell areas, and interface edge lengths, and assuming that cell areas depend on the pressure balance generated by interfacial tension and spring tension, we generate an effective bulk model (see SM Sec. E):
\begin{equation}
    \tilde{e}_{tot} = (\tilde{\langle a \rangle} - 1)^{2} + \tilde{k}_{p} (\tilde{\langle p \rangle} - \tilde{p}_{0})^{2}.
\label{eq_reformulated_energy_func}
\end{equation}{}
The effective model parameters are denoted by tildes and are functions of the original model parameters, including $\gamma$, and the number of tissue cells, $N_{cells}$. A plot of the effective parameter $\tilde{p}_{0}$ is shown as a function of $\gamma$ in Fig. 2 (c). Given the form of this effective energy functional, we predict that a system with a finite boundary and non-zero interfacial line tension will behave as a bulk vertex model with effective parameters, $\tilde{p}_{0}$ and $\tilde{k}_{p}$. Although the critical $p_{0}$, $p_{0, c}$, in the bulk vertex model has been shown to depend on temperature and simulation protocol, the zero-temperature value of $p_{0, c}=3.81$ appears to represent a lower-bound, and at our simulation temperature we expect a transition from fluid-like to solid-like behavior at $\tilde{p}_{0}(p_{0}, \gamma, N_{cells}) \approx 3.81$ (see Refs.  \cite{Bi_2016, Sussman_2018, Wang_2019} and SM Sec. A). Indeed we find that the value of $\gamma$ at which $\tilde{p}_{0}=3.81$ corresponds well to the point at which $D_{eff}$ becomes non-zero, as noted by the dashed lines in Fig. \ref{fig_case_01_panel} (b) and (c). This mapping explains our observation of fluid-like behavior with increasing interfacial tension; even for $p_{0} < 3.81$, we expect fluid-like behavior for associated $\tilde{p}_{0} > 3.81$. Moreover, since the relationship between $p_{0}$ and mean observed cell shape, $\langle s \rangle$, has been shown for the bulk vertex model, our mapping also predicts the behavior of $\langle s \rangle$ with increasing $\gamma$, or, equivalently, $\tilde{p}_{0}$. This relationship, along with that for a bulk vertex model alone with varying $p_{0}$, is illustrated in Fig. \ref{fig_case_01_panel} (d). 

Although these results focus on the limit of floppy ECM, adding tension to the network, by decreasing $l_{0}$, simply shifts the mean cell area, as shown in Fig. \ref{fig_full_model} (d). By identifying another mean-field approximation relating the mean spring length to the mean cell area, our effective model parameters become functions of both $\gamma$ and $l_{0}$ (see SM Sec. E). 

\begin{figure}
    \centering
    \includegraphics[width=0.45\textwidth]{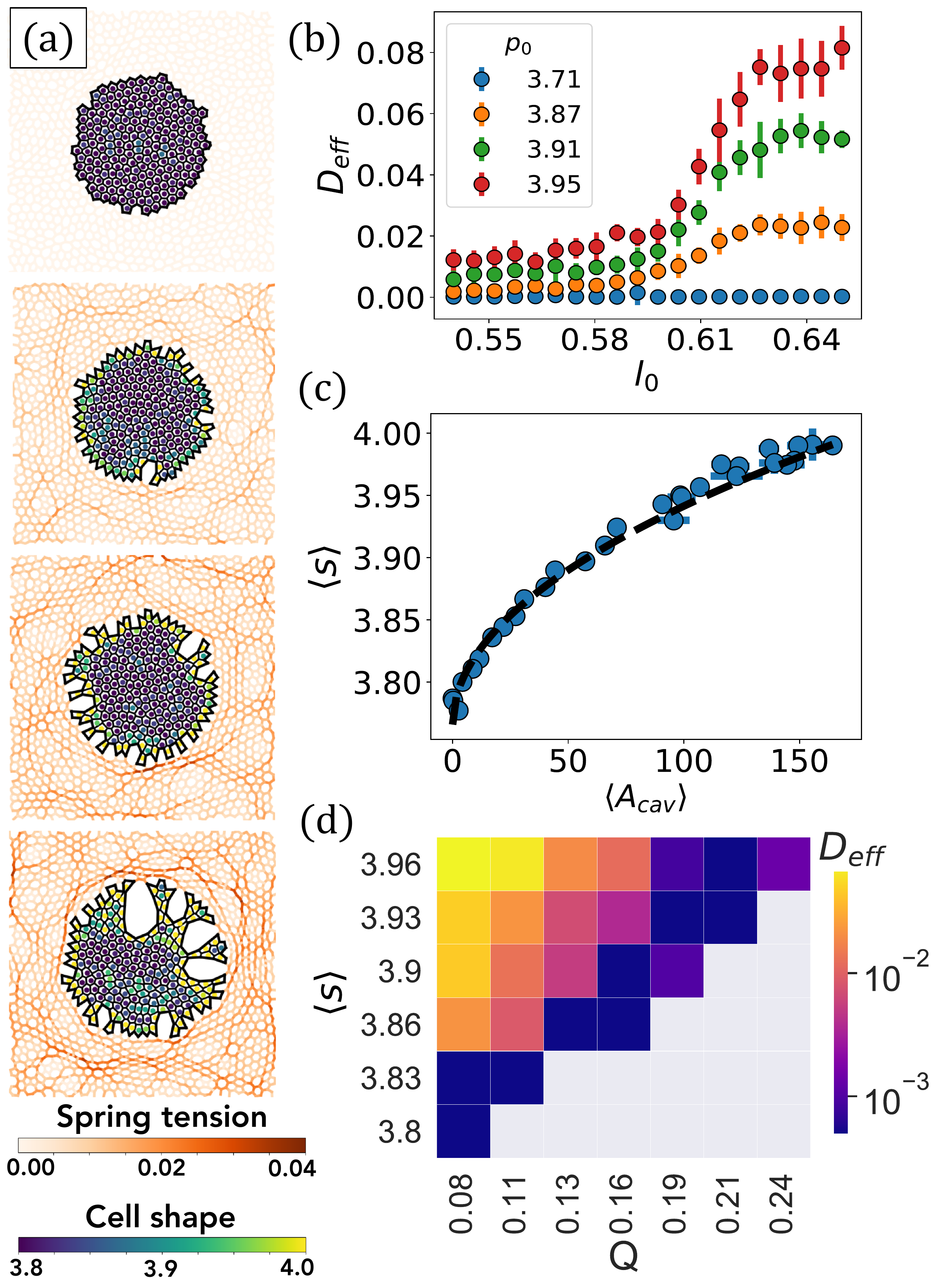}
    \caption{\textit{Irregular tissue regime.} (a) Minimium-energy configurations for tissues with $p_{0}=3.70$ and increasing ECM tension (decreasing $l_{0}$) for $\gamma=0$. From top to bottom, $l_{0}$ = 0.63, 0.61, 0.60, and 0.59. Each spring is colored by its tension ($T_{\langle ij \rangle} = k_{sp}(l_{\langle ij \rangle} - l_{0})$), and each cell center is colored by its shape, $s = p/\sqrt{a}$, where $p$ and $a$ are the cell's perimeter and area. (b) The effective diffusion coefficient, $D_{eff}$, vs. $l_{0}$ of the surrounding spring network, for a range of $p_{0}$ values for the tissue cells. (c) The mean cell shape, $\langle s \rangle$, vs. the total cavity area, $A_{cav}$, for a tissue of $p_{0}=3.70$ (points), and a fit of the form in Eq. \ref{eq_case_02_cell_perim_final} (dashed). (d) A heat map colored by $D_{eff}$, for varying mean cell shape, $\langle s \rangle$, and cell-cell alignment, $Q$. }
    \label{fig_case_02_panel}
\end{figure}

\noindent {\it \textbf{Irregular tissue regime.}} At $\gamma_{c}$, the tissue boundary becomes unstable, and for $\gamma < \gamma_{c}$, we observe irregular tissue boundaries facilitated by the opening of cavities-- empty spaces along which cells are not coupled to ECM springs. To study this regime, we first explore the limit of increasing tension in the spring network with no tissue interfacial tension. In this limit, there is no cost to opening gaps that increase the length of the spheroid-ECM interface, and so, as $l_{0}$ decreases below the critical threshold and the tension in the ECM increases, gaps open. Minimum energy configurations are illustrated in Fig. \ref{fig_case_02_panel} (a). As the spring network tension increases, the tissue boundary becomes increasingly irregular and the average cavity size grows.

To study the phase of the tissue, we again run low-temperature simulations and extract the effective diffusion constant, $D_{eff}$, in this case for a range of $l_{0}$ values. For a fixed $p_{0}$ and decreasing $l_{0}$, $D_{eff}$ decreases. Examples of this are shown in Fig. \ref{fig_case_02_panel} (b). To determine if the observed cell shape is correlated with the decrease in diffusivity, we ignore fluctuations in cell perimeters, areas, and spring lengths and treat each cell as bordering a cavity and being stretched along the cavity's circular circumference. We also assume the cell areas do not increase, which is substantiated by numerics. As each cell becomes highly elongated in the tangential direction and compressed in the radial direction, the relationship between the cell perimeter and the radius of the cavity approaches a linear form (see SM Sec. G). Therefore, the perimeter of the cell scales with the square root of the area of the cavity, and, since we assume the cell area is fixed, so does the cell shape. Assuming the total cavity area, $A_{cav}$, is a multiple of the single cavity area, we obtain 
\begin{equation}
   \langle s \rangle = B_{1} \sqrt{A_{cav}} + B_{2}.
\label{eq_case_02_cell_perim_final}
\end{equation}{}
In Fig. \ref{fig_case_02_panel} (c), simulation results for $p_{0}=3.70$ and a fit of the form of Eq. \ref{eq_case_02_cell_perim_final} are shown. Since we anticipate bulk tissue behavior when $A_{cav}=0$, we expect $B_{2} \approx 3.81$ (see SM Sec. A). We find that $B_{1}=0.0176(7)$ and $B_{2}=3.766(3)$. These results, along with the observed decrease in fluidity, indicate that, unlike the {\it compact tissue regime}, a mean cell shape above $p_{0,c}$ does not necessarily correspond to fluid-like tissue. 

Prior work has shown that bulk tissue under anisotropic strain can rigidify due to cell-cell alignment for $\langle s \rangle > p_{0, c}$~\cite{Wang_2019}. To determine if this phenomenon is relevant here, we quantify cell-cell alignment with the parameter $Q$~\cite{Wang_2019}, which is similar to a nematic order parameter except that it also contains information about the shape of the cells (see SM Sec. H). As illustrated in Fig. \ref{fig_case_02_panel} (d), the decrease in $D_{eff}$ is indeed correlated with an increase in cell shape and an increase in cell-cell alignment. Zero-temperature energy barrier measurements also support this finding (see SM Sec. I). Note that although there is a clear decrease in $D_{eff}$ as $l_{0}$ decreases, a difference remains across $p_{0}$ values. This suggests that the rigidification may not necessarily be a bulk phenomenon but become enhanced toward the periphery of the spheroid. Energy barrier measurements as a function of distance from the periphery suggest this. Finite-size and temperature studies are in SM Secs. J and K. 

\noindent {\it \textbf{Discussion.}} 
We have generalized the bulk vertex model for cellularized tissue to include a coupling to a surrounding spring network representing the ECM. By studying this bi-material, we gain insight into the competition between cell-cell and cell-environment interactions. Our findings represent the essential first step in understanding a model that can now easily be augmented with specific features. When the tissue remains compact, isotropic compression-- driven by increasing interfacial tension-- leads to increased cell shapes and fluidization of the tissue, as predicted by a mean-field mapping to a bulk tissue. Moreover, a transition from a compact tissue geometry to an irregular one happens suddenly, due to a generic instability in the balance of negative pressure induced by the surrounding network and positive pressure caused by the interfacial tension. Once gaps open between the tissue and the ECM, the rheology of the tissue depends both on cell shape and alignment, as in the case of tissues under shear deformations \cite{Wang_2019}. We note that the tissue deformation in our system is non-uniform, due to the system's disorder and heterogeneity, resulting in heterogeneity in alignment and shape among the tissue cells. It has been shown that heterogeneity itself can promote rigidity in tissues \cite{Li_2019}, which perhaps also contributes to the induced rigidity of the system. 

Our results suggest that although stiff networks have been shown to induce tissue breakup and cell migration, this may not mean they induce tissue fluidity. In fact, stiff networks may induce stiffer tissues, due to emergent cellular alignment, despite irregular tumor boundaries and elongated cell shapes. In other words, describing cancer invasion as cellular unjamming may be over-simplifying the phenomenon, as some experiments have recently indicated~\cite{Kang2020}. The pre-invasion stage of the tumor may involve solidification due to cellular alignment, ultimately facilitating cellular streaming upon enhanced cellular activity, with the tissue becoming an active fluid, flowing with respect to the ECM. Measurements of cellular alignment are needed to test this idea. Experiments searching for compression-induced fluidization in compact tumors or looking for cavity formation at the tumor-ECM interface would also test the model, since, given the generic nature of our arguments, we expect the phenomena to survive in three-dimensions. Finally, our model lays the groundwork for understanding more complex scenarios. For example, the ability for cells to form or degrade cell-ECM contacts, cell self-propulsion, ECM reorganization, and feedback between cell-cell and cell-ECM interactions are all fascinating directions for future study.

We acknowledge NSF-PoLS 1607416 for financial support. We would like to thank Daniel Sussman and Preeti Sahu for stimulating discussions and guidance on use of the cellGPU codebase. We acknowledge Mingming Wu for feedback on early results. 


%

\end{document}


\section{Supplementary Material}

\setcounter{secnumdepth}{3}
\setcounter{equation}{0}
\setcounter{figure}{0}  
\setcounter{table}{0}

\makeatletter 
\renewcommand{\thefigure}{S\@arabic\c@figure}
\makeatother

\makeatletter 
\renewcommand{\thetable}{S\@Roman\c@table}
\makeatother

\makeatletter 
\renewcommand{\theequation}{S\@arabic\c@equation}
\makeatother

\renewcommand{\bibnumfmt}[1]{[S#1]}
\renewcommand{\citenumfont}[1]{S#1}

\subsection{Vertex models for tissues} \label{sec_vertex_model_review}

Vertex models of epithelial tissues represent groups of cells as space-tiling polygons. A configuration is defined by a set of vertex positions and, for each vertex, a set of neighbor vertices. The connections between vertices and their neighbors form the edges of the polygonal tissue cells. The degrees of freedom in the system are the coordinates of the vertices, which evolve in a simulation, enabling changes in the shapes of the tissue cells over time. For a 2D vertex model, a typical energy functional is
\begin{multline}
    E_{VM} = \sum_{\alpha=1}^{N_{cells}} K_{A, \alpha}(A_{\alpha} - A_{0, \alpha})^{2} \\ + \sum_{\alpha=1}^{N_{cells}} K_{P,\alpha}P_{\alpha}^{2}
    + \frac{1}{2} \sum_{\langle ij \rangle=1}^{N_{edges}} \Lambda_{\langle ij \rangle} l_{\langle ij \rangle}
\label{eq_vertex_model_00}
\end{multline}{}
where $A_{\alpha}$ and $P_{\alpha}$ are the area and perimeter of cell $\alpha$, $l_{\langle ij \rangle}$ is the length of edge $\langle ij \rangle$, and $A_{0, \alpha}$, $K_{A, \alpha}$, $K_{P,\alpha}$ and $\Lambda_{\langle ij \rangle}$ are the target area, area stiffness, contractility strength, and line tension of cell $\alpha$, respectively. 
The terms in the first sum in Eq. \ref{eq_vertex_model_00} capture the 2D incompressibility of the cells. The second term captures active contraction in acto-myosin cables, and the third term captures cell-cell adhesion and cortical tension generated by acto-myosin \cite{Farhadifar_2007, Bi_2015}. If we assume that all constants are the same for each cell and edge, we can pull the constants outside the sums, and write $A_{0, \alpha}$ as simply $A_{0}$. The sum in the last term can now be written in terms of cell perimeters, since it is a sum over all cell edge lengths. We then have
\begin{equation}
    E_{VM} = K_{A} \sum_{\alpha=1}^{N_{cells}} (A_{\alpha} - A_{0})^{2} + K_{P} \sum_{\alpha=1}^{N_{cells}} P_{\alpha}^{2} + \Lambda \sum_{\alpha=1}^{N_{cells}}  P_{\alpha}.
\label{eq_vertex_model_01}
\end{equation}{}
Now, for each cell, we combine the last two terms and, dropping constants that simply shift the total energy, we obtain
\begin{equation}
    \tilde{E}_{VM} = K_{A} \sum_{\alpha=1}^{N_{cells}} (A_{\alpha} - A_{0})^{2} + K_{P} \sum_{\alpha=1}^{N_{cells}} (P_{\alpha} - P_{0})^{2},
\label{eq_vertex_model_02}
\end{equation}{}
where $P_{0}=-\frac{\Lambda}{2 K_{P}}$ is the target cell perimeter, common to all cells. Non-dimensionalizing Eq. \ref{eq_vertex_model_02}, using $K_{A}A_{0}^{2}$ as our unit of energy, reduces the number of free parameters to leave us with
\begin{equation}
    e_{VM} = \sum_{\alpha=1}^{N_{cells}} (a_{\alpha} - 1)^{2} + k_{p} \sum_{\alpha=1}^{N_{cells}} (p_{\alpha} - p_{0})^{2},
\label{eq_vertex_model_03}
\end{equation}{}
where we use lower-case letters to indicate dimensionless parameters (and drop the tilde on the left-hand side), $k_{p}=\frac{K_{P}}{K_{A} A_{0}}$ and $p_{0}=\frac{P_{0}}{\sqrt{A_{0}}}$.

The parameter $p_{0}$ is the dimensionless target perimeter or the target, or preferred, ``shape index'' of the cells. This follows from defining the observed ``shape'' of a cell as $s = P/\sqrt{A}$, where $P$ and $A$ are the actual perimeter and area of the cell, respectively. This quantity, $s$, captures how circular a cell is, with lower values being more circular and higher values being less circular (which in many cases means more elliptical or elongated). Since $p_{0}$ is defined as $\frac{P_{0}}{\sqrt{A_{0}}}$, it therefore acts as a ``target'' shape for the cells. 

It has been shown that the preferred shape, $p_{0}$, can act as a tuning parameter for the bulk 2D vertex model. One way to measure the phase of a tissue is to calculate an effective diffusion coefficient, $D_{eff}$, for the cells, which is approximately zero for $p_{0} < p_{0, c}$ and greater than zero for $p_{0} > p_{0, c}$ \cite{Bi_2015}. In order for cells to diffuse in a system with no gaps between them (a ``confluent'' tissue), they must undergo T1 transitions, which allow them to exchange neighbors even with the confluency constraint (see Sec. \ref{sec_t1_transitions} for more information).

This transition also manifests in the relationship between the observed mean cell shape, $\langle s \rangle$, and preferred cell shape, $p_{0}$. In the 2D, disordered vertex model, for $p_{0} < p_{0, c}$, cells become frustrated as they cannot achieve their preferred shape, while for $p_{0} > p_{0, c}$, $\langle s \rangle = p_{0}$. This comes from the fact that for a fixed area, the perimeter of a polygon has a lower bound, but no upper bound.

Previous work has shown that the value of $p_{0, c}$ depends on temperature (the magnitude of thermal fluctuations) and simulation protocol, and the often-cited $p_{0, c}=3.81$ appears to be a lower-bound on the transition point \cite{Bi_2016, Sussman_2018, Wang_2019}. Fig. S3 of Ref. \cite{Wang_2019} and Fig. 5 of Ref. \cite{Bi_2016} are particularly illuminating. Our Brownian dynamics simulations are run at temperatures considered low according to these studies (see Sec. \ref{sec_brownian_dynamics_methods}), and therefore expect and find the transition point to be close to $p_{0}=3.81$, but do not expect it to match exactly. 

\subsection{Spring network models} 
\label{sec_spring_network_model_review}

A spring network model is a system, like the vertex model for tissues, consisting of vertices and vertex-vertex neighbors, with the positions of the vertices as the degrees of freedom. A simple Hamiltonian for a spring network model assigns a spring-like energy cost to a deviation of a vertex-vertex neighbor distance, $L_{\langle ij \rangle}$, from some ``preferred'' or ``target'' distance, $L_{0}$:
\begin{equation}
    E_{SNM} = \frac{1}{2} \sum_{sp \langle ij \rangle }^{N_{sp}} K_{sp \langle ij \rangle} (L_{sp \langle ij \rangle} - L_{0})^{2}.
\label{eq_spring_network_model_00}
\end{equation}
This model does not include bending penalties that are typically included in fiber network models. However, recent work suggests that the mechanics of such networks is largely controlled by the critical point in simple spring network models \cite{Shivers_2019}.

This model does not have particles or cells that can transition from a diffusive phase to a caged phase. One can, however, quantify the difference between a fluid and solid phase by considering the response to an imposed strain. A floppy (fluid-like) system has no mechanical response to strain, while a solid system resists strain in the form of an energy cost. Spring networks, even without more physically-realistic bending energies, also exhibit this behavior. 

The method of Maxwell constraint counting demonstrates that determining whether a network will be floppy or rigid at small strain is equivalent to determining whether there are more degrees of freedom or constraints in the network \cite{Maxwell_1864}. If the number of degrees of freedom is greater than the number of constraints, the system is referred to as ``under-constrained,'' and the system is floppy. However, even a floppy network can become rigid above a critical strain value, at which point the network becomes geometrically frustrated. In other words, due to the connectivity of the network, there is a point at which the new imposed deformation of the network cannot be accommodated without an energy cost, and the system is rigid \cite{Sharma_2016, Merkel_2019}.

\begin{figure}
    \centering
    \includegraphics[width=0.45\textwidth]{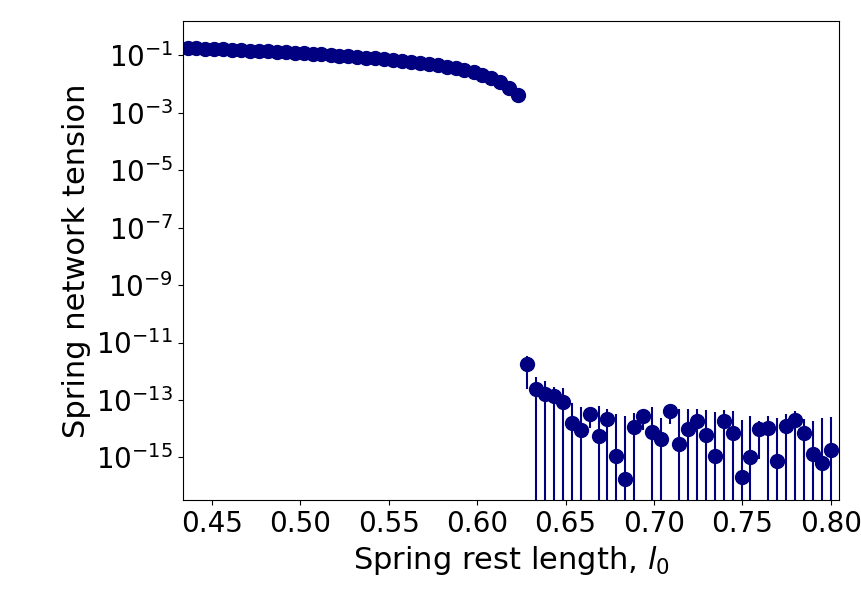}
    \caption{The average spring network tension as a function of spring rest length, $l_{0}$, for a system of 500 polygons, where each edge is assigned a spring energy. In this example, the system is purely a spring network, with no embedded tissue cells. At a critical value of $l_0$, the tension in the network increases by about 10 orders of magnitude.}
    \label{fig_l0_vs_tension_pure_spring_log_scale}
\end{figure}

In order to impose a strain on a spring network, one can externally deform the network's boundaries. In simulations, this is often done by shearing or uniformly expanding the boundaries. Alternatively, for fixed boundaries, one can impose a new $L_{0}$ value for the springs. The effect of this is similar to that of imposing an external uniform expansion of the network, without having to introduce a new strain parameter. Indeed, we see that a simple two-body spring network will transition from a floppy to rigid phase as a function of $L_{0}$, as shown in Fig. \ref{fig_l0_vs_tension_pure_spring_log_scale}. 

It is not a coincidence that both the spring network and the vertex model have a rigid-to-floppy transition that is tuned by a preferred length parameter ($L_{0}$ and $p_{0}$, respectively). In fact, it has been shown that these parameters are deeply related, and that the transitions in both models are of the same universality class \cite{Merkel_2019}.

{}

\subsection{T1 transitions} 
\label{sec_t1_transitions}

\begin{figure}
    \centering
    \includegraphics[width=0.3\textwidth]{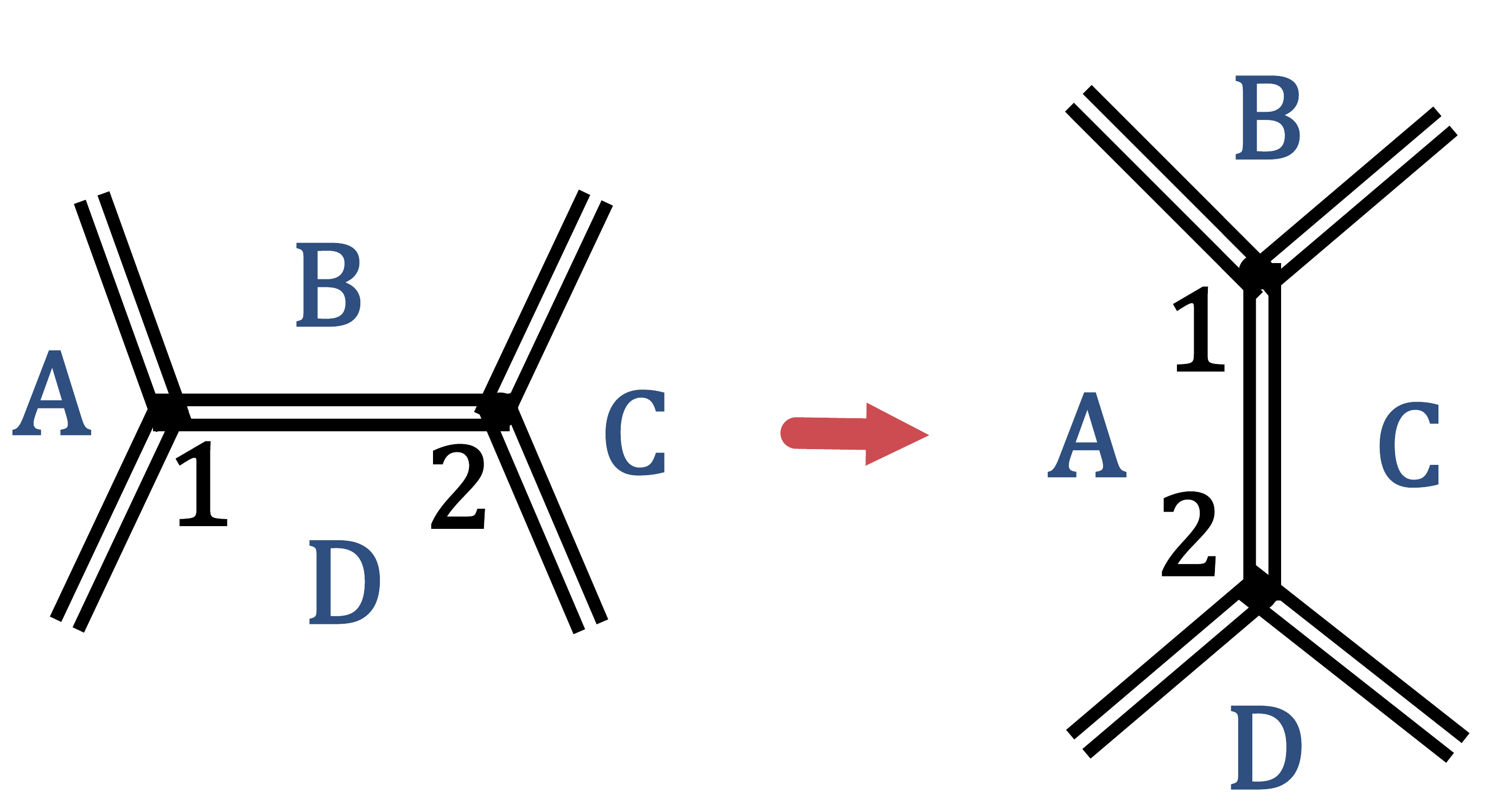}
    \caption{An illustration of a T1 transition in bulk tissue. Tissue cells are labeled with letters and vertices defining the edge undergoing the T1 are labeled with numbers. Edge 1-2 shrinks to a point (not shown) and then elongates in the direction perpendicular to its initial configuration. Notice that cells B and D are no longer neighbors after the T1, while cells A and C have become neighbors. This demonstrates that cells can exchange neighbors and therefore move through a tissue, even when there are no spaces between cells.}
    \label{fig_generic_t1}
\end{figure}

In a confluent tissue, there are no gaps between cells. In a fluid-like state, however, cells move through the tissue, flowing like a liquid under external forces. This is possible due to a process called a T1 transition. During a T1 transition, an edge in the network shrinks to a point and a new edge then expands from that point, in the direction perpendicular to the original edge (see Fig. \ref{fig_generic_t1}). In this way, cells can exchange neighbors and translate, while staying in the plane of the tissue and without needing empty space to move into.

{}

\begin{figure}
    \centering
    \includegraphics[width=0.45\textwidth]{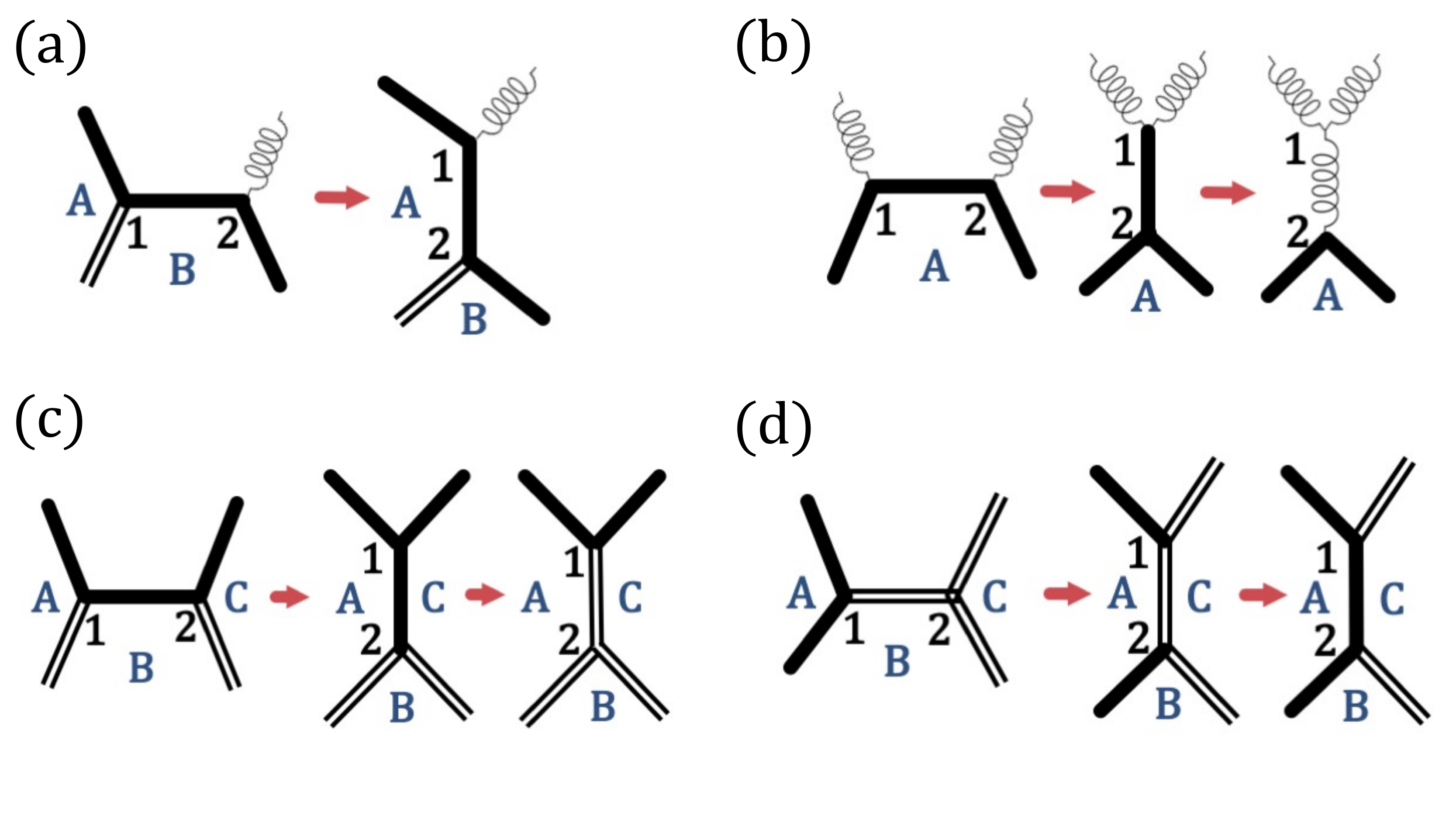}
    \caption{Illustrations of four special cases of T1 transitions involving edges along the tissue-ECM boundary. The vertices of the edge undergoing a T1 are labeled 1 and 2 in each case. Tissue cells are labeled with capital letters, while empty space in the ECM network is unlabeled. Spring edges are represented using a coil icon, edges on the interface are represented with solid lines, and edges between two tissue cells are represented with double lines.}
    \label{fig_T1_cases}
\end{figure}

In our bi-material, any edge between two tissue-like vertices is allowed to undergo a T1. This refers to edges in the tissue bulk. In the spring network, the opposite is true: no T1 transitions are allowed along spring edges. Edges that are between two interface vertices or an interface vertex and a tissue vertex, however, have to be considered separately. In describing the cases of interest, we will refer to Fig. \ref{fig_T1_cases}, where interface edges are represented as thick lines, tissue edges are represented as double lines, and spring edges are represented as a spring icons. Tissue cells are labeled with capital letters and the edge that is undergoing the T1 has vertices labeled 1 and 2. Polygons that are not tissue cells but are instead empty space in the spring network are given no label.

{}

The first case to consider is illustrated in Fig. \ref{fig_T1_cases} (a). In this case, vertex 1 is also a neighbor to a spring-type vertex and another interface-type vertex, and vertex 2 is also a neighbor to a tissue-type vertex and another interface-type vertex. In this case, a T1 simply pushes the cell that shared edge 1-2, cell B, along the boundary. The edge remains an interface-type edge. The reverse of this case (moving from the right-hand-side of the figure to the left) is also allowed and the procedure is identical to the forward case, except that the vertex labels 1 and 2 are switched.

The second case, Fig. \ref{fig_T1_cases} (b), concerns an interface edge in which both vertices 1 and 2 have a spring-type neighbor vertex and another interface-type neighbor vertex. Here, a T1 pushes the cell that shares edge 1-2, cell A, away from the boundary, into the bulk. Edge 1-2 is then shared by two spring-network-type polygons, and should be a spring-type edge. Therefore it automatically gets relabeled as a spring. The reverse of this is not allowed, as T1s are not allowed along spring-type edges. In other words, an interface edge that goes through this T1 and becomes a spring cannot revert back to an interface edge. Note that an interface edge is lost in this T1.  

The third case, Fig. \ref{fig_T1_cases} (c),  is analogous to the second, except that instead of springs, vertices 1 and 2 each has a tissue-type neighbor in addition to their interface-type neighbor. Here the T1 also forces the adjacent cell, cell B, into the bulk, but the edge now changes from an interface-type to a tissue-type due to its new cell neighbors. The reverse of this procedure is allowed and is described in case 4.  Case 3 results in a loss of an interface edge, while the reverse leads to interface edge creation. 

Case four, Fig. \ref{fig_T1_cases} (d), concerns a T1 between an interface-type vertex and a tissue-type vertex. After a T1, this originally tissue-type edge is now on the boundary, as a bulk cell, cell C, has moved from the bulk to the boundary. Therefore this edge gets updated to an interface-type edge, i.e. a new interface edge emerges. The reverse of this process is described in case three.

While both interface edge creation and removal are allowed, interface edge creation via the cavitation instability is much more common than interface edge removal.

\subsection{Simulation details: algorithms and parameters}
\label{sec_simulation_algorithms}

To numerically study the bi-material, we implement a modified version of the open-source cellGPU software \cite{Sussman_2017}. This molecular dynamics software is written specifically for vertex-like models of tissues and has the ability to run on GPUs for especially efficient performance, although it can also be run on CPUs. 

As is true for most molecular dynamics software, cellGPU uses a force field-plus-equation of motion approach, where one can essentially ``mix-and-match'' potential energy functions and equations of motion. Therefore, adding the force field (potential energy and corresponding force functions) for our coupled tissue-ECM model, and all necessary data structures not already included (like spring moduli, interface edge moduli, etc.), allows us to then interface with the already-existing algorithms (described below) for zero-temperature energy minimization and low-temperature Brownian dynamics.

For all simulations, we begin with $N_{total}=1000$ polygons tiling a periodic box. We then define a circle whose center is the box center and whose area is 20\% of the total box area. Any polygon whose center is within this circle is labeled a tissue-type cell, and any polygon outside this circle is labeled an ECM-type polygon. The edges shared by different types of polygons are labeled interface-type. The number of tissue-type cells varies slightly from configuration to configuration, but is always very close to $1000 \times 0.20 = 200$ cells. We have studied additional ratios of the number of cells to springs (see Sec. J). 

\subsubsection{Energy minimization simulation procedure}
\label{sec_energy_minimization_methods}

All results except those for the effective diffusion coefficient are obtained using energy minimization simulations. Specifically, we use the FIRE minimization algorithm \cite{Bitzek_2006}, where the equation of motion is given by
\begin{equation*}
    \vec{v}{\,}' = \vec{v} + \vec{F}\Delta t + \alpha (|\vec{v}|\hat{F} - \vec{v}).
\end{equation*}
For a given particle's (vertex's) current position and velocity, the algorithm calculates the current forces on the particle ($\vec{F} = -\nabla E(\vec{x})$) and the power, $P=\vec{F} \cdot \vec{v}$. If $P$ is negative, it means that the forces would move us back ``uphill,'' so we decrease our timestep, $dt$, by some fraction, $f_{dec}$, such that $dt ^{\prime} = dt \times f_{dec}$, set the velocity to zero, and set $\alpha$ back to its initial value. If instead $P$ is positive, and the number of timesteps since $P$ was negative is greater than some chosen minimum value, $n_{min}$, we increase our timestep by some fraction, $f_{inc}$ (unless we are already at our maximum allowed timestep, $dt_{max}$, in which case we do not change the timestep) and decrease $\alpha$ by some fraction $f_{\alpha}$. The simulation ends when the maximum force in the system is less than or equal to the chosen force cut-off, $F_{cut-off}$. Typical values for these parameters used in our simulations are given in table \ref{tbl_energy_minimization_params}.

\begin{table}{}
\begin{center}
\renewcommand{\arraystretch}{1.5} 
\begin{tabular}{c|c}
   Parameter & Value \\
   \hline
   \hline
    $dt_{start}$ & $ 1 \times 10^{-4}$ \\
    $\alpha_{start}$ & $0.99$ \\
    $dt_{max}$ & $1000 \times dt_{start}$ \\
    $f_{inc}$ & $1.1$ \\
    $f_{dec}$ & $0.95$ \\
    $\alpha_{dec}$ & $0.9$ \\
    $n_{min}$ & $4$ \\
    $F_{cut-off}$ & $1 \times 10^{-8}$ \\
\end{tabular}{}
\caption{\label{tbl_energy_minimization_params} Typical parameter values used during FIRE minimization simulations.}
\end{center}{}
\end{table}

\subsubsection{Brownian dynamics simulation procedure and calculation of mean squared displacements and effective diffusion coefficients}
\label{sec_brownian_dynamics_methods}

In cases where we are not just interested in the static energy-minimum of a system, we use Brownian dynamics (over-damped Langevin dynamics) to simulate the motion of vertices over time. This equation of motion is given by
\begin{equation*}
    \vec{x}{\,}' = \vec{x} + \mu \vec{F} dt + \sqrt{2\mu k_{B}T dt}\vec{R}(t)
\end{equation*}
where $\vec{F} = -\nabla E(\vec{x})$, $\mu$ is an inverse damping coefficient, $T$ is the temperature, and $\vec{R}(t)$ is a vector of delta-correlated random numbers with a Gaussian distribution and zero mean. Typical values used for the parameters in our simulations are given in table \ref{tbl_brownian_dynamics_params}.

\begin{table}{}
\begin{center}
\renewcommand{\arraystretch}{1.5} 
\begin{tabular}{c|c}
   Parameter & Value \\
   \hline
   \hline
    $\mu$ & $1.0$ \\
    $T$ & $1 \times 10^{-5}$ \\
    $dt$ & $0.01$ \\
    $k_{B}$ & $1.0$
\end{tabular}{}
\caption{\label{tbl_brownian_dynamics_params} Typical parameter values used during Brownian dynamics simulations.}
\end{center}{}
\end{table}

For each simulation, we run at least $1 \times 10^{6}$ initial timesteps at which point the system has come to a steady state. We then run for an additional $1 \times 10^{8}$ timesteps during which data on the state of the system is periodically saved. We use this data to calculate the mean squared displacement (MSD) of the tissue cell centers. This is defined as
\begin{equation*}
    \text{MSD}(\tau) = \langle (\vec{r}(t + \tau) - \vec{r}(t))^{2} \rangle
\end{equation*}{}
where $t$ is the time of some reference state, in units of our timestep. This average, for a single simulation, is over cells and reference times. The MSDs for all simulations are then averaged and this mean and standard deviation is used as our result. 

\begin{figure}
    \centering
    \includegraphics[width=0.45\textwidth]{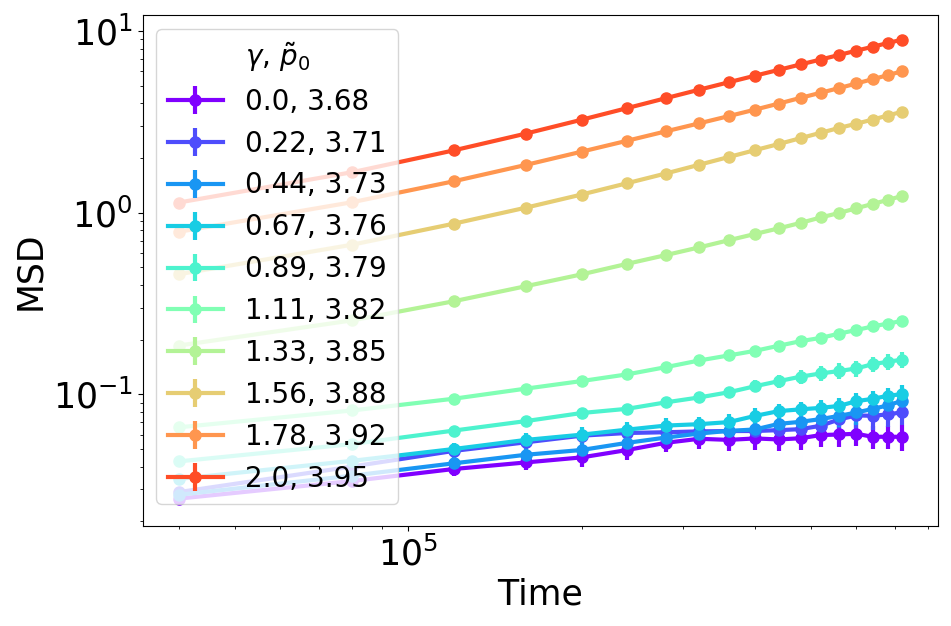}
    \caption{Mean squared displacement (MSD) for fixed $p_{0}=3.68$ and varying interfacial tension, $\gamma$, values (and therefore varying $\tilde{p}_{0}$ values), vs. simulation time. The final data points are used to calculate the effective diffusion coefficient for each $\tilde{p}_{0}$, which are reported in the main text.}
    \label{fig_msd_vs_time}
\end{figure}{}

The effective diffusion coefficient, $D_{eff}$, for the cells is defined as
\begin{equation*}
    D_{eff} = \frac{D_{s}}{T} = \frac{1}{T} \lim_{\tau \rightarrow \infty} \frac{\text{MSD}(\tau)}{4 \tau}.
\end{equation*}{} 
where, instead of taking $\tau$ to infinity, we use $\tau = 4 \times 10^{5}$ (in units of the simulation timestep length) as our long-time estimate. 

An example of a set of MSDs are shown in Fig. \ref{fig_msd_vs_time}. The $D_{eff}$ values calculated from this particular example are used to produce Fig. 2 (b) in the main text.

\subsection{Analytical model: compact tissue regime}
\label{sec_regime_1_math_details}

To model the full \textit{compact tissue regime}, our goal is to express the sum over interfacial edges and sum over springs in terms of the mean cell area, in order to construct an effective theory of only cell areas and perimeters. To do so, we first simplify the energy function (Eq. 1 of the main text) by ignoring fluctuations from the mean cell area, $\langle a \rangle$, mean cell perimeter, $\langle p \rangle$, mean interface edge length, $\langle l_{int} \rangle$, and mean spring length, $\langle l_{sp} \rangle$. This gives 
\begin{multline}
    e_{total} = N_{cells}(\langle a \rangle - 1)^{2} + k_{p} N_{cells} (\langle p \rangle - p_{0})^{2} 
    \\+ \gamma N_{int} \langle l_{int} \rangle 
    \\+ \frac{1}{2} k_{sp} N_{sp} (\langle l_{sp} \rangle - l_{0})^2,
\label{eq_nondim_energy_total_high_gamma_01}
\end{multline}
where $N_{sp}$ is the total number of springs, $N_{cells}$ is the total number of cells, and $N_{int}$ is the total number of interface edges.

The tissue's total perimeter is equal to the sum over the lengths of the interface edges. Ignoring fluctuations gives
\begin{equation}
    P_{tissue} = \sum_{int \langle ij \rangle} l_{int \langle ij \rangle} = N_{int}\langle l_{int} \rangle.
\end{equation}
At the same time, since the tissue in this regime is highly circular, the perimeter is also related to the tissue radius, $R_{tissue}$:
\begin{equation}
    P_{tissue}= 2 \pi R_{tissue}.
\end{equation}{}
Therefore, the $R_{tissue}$ is given by
\begin{equation}
    R_{tissue} = \frac{N_{int}\langle l_{int} \rangle}{2 \pi}.
\label{eq_tissue_radius}
\end{equation}

Since the tissue is circular, its area is also a function of its radius, and we can write the total tissue area in terms of Eq. \ref{eq_tissue_radius}:
\begin{equation}
    A_{tissue} = \pi R_{tissue}^{2} = \frac{N_{int}^{2} \langle l_{int} \rangle^{2}}{4 \pi}.
\label{eq_tissue_area_01}
\end{equation}{}
The area of the tissue must also be the sum of the constituent cell areas, which we approximate as each equal to the mean cell area. This gives
\begin{equation}
    A_{tissue} = \sum_{cell, \alpha} a_{\alpha} = N_{cells}\langle a \rangle.
\label{eq_tissue_area_02}
\end{equation}{}
Combining Eq. \ref{eq_tissue_area_01} with Eq. \ref{eq_tissue_area_02} gives the average cell area as
\begin{equation} 
    \langle a \rangle = \frac{N_{int}^{2} \langle l_{int} \rangle^{2}}{4 \pi N_{cells}}
\label{eq_mean_cell_area_no_fluc}
\end{equation}
Rearranging Eq. \ref{eq_mean_cell_area_no_fluc}, we write the mean interface edge length as a function of the mean cell area:
\begin{equation}
    \langle l_{int} \rangle = \sqrt{\frac{4 \pi N_{cells}}{N_{int}^{2}}} \sqrt{\langle a \rangle}
\label{eq_mean_interface_edge_length}
\end{equation}{}

Now, to write $\langle l_{sp} \rangle$ in terms of $\langle a \rangle$, we approximate the spring network representing the ECM as a regular, hexagonal lattice. In that case, the number of hexagons in the network is
\begin{equation}
    N_{hex} = \frac{1}{6}(N_{springs})(2 \frac{hexagons}{spring}) = \frac{1}{3} N_{springs},
\label{eq_supp_number_hexagons}
\end{equation}
since each spring is shared by two hexagons. The factor of $1/6$ accounts for over-counting. The total area taken up by the spring network is the sum of the areas of the hexagons:
\begin{equation}
    A_{sp.net} = N_{hex} A_{hex} = \frac{1}{3} N_{sp}(\frac{3\sqrt{3}}{2} \langle l_{sp} \rangle^{2}),
\label{eq_supp_area_spring_net_01}
\end{equation}
where $A_{hex}$ is the area of a regular hexagon with edge length equal to $\langle l_{sp} \rangle$. The total area taken up by the spring network is also equal to the total box area minus the total tissue area, which is itself a sum of the cell areas. Ignoring fluctuations in cell areas as before, this gives
\begin{equation}
    A_{sp.net} = A_{box} - A_{tissue} = A_{box} - N_{cells} \langle a \rangle.
\label{eq_supp_area_spring_net_02}
\end{equation}
Combining Eqs. \ref{eq_supp_area_spring_net_01} and \ref{eq_supp_area_spring_net_02} and solving for $\langle l_{sp} \rangle$ gives
\begin{equation}
    \langle l_{sp} \rangle = \sqrt{\frac{2}{\sqrt{3}}} \sqrt{\frac{A_{box} - N_{cells}\langle a \rangle}{N_{sp}}}.
\label{eq_supp_lspring_01}
\end{equation}{}

Now let $\langle a \rangle = 1 + \epsilon$, where $| \epsilon | < 1$ and $\epsilon$ can be positive or negative. Then Eq. \ref{eq_mean_interface_edge_length} becomes
\begin{equation}
    \langle l_{int} \rangle = \sqrt{\frac{4 \pi N_{cells}}{N_{int}^{2}}} \sqrt{1 + \epsilon}.
\label{eq_mean_interface_edge_length_epsilon}
\end{equation}{}
Taylor expanding the square root about $\epsilon=0$ yields
\begin{equation}
    \langle l_{int} \rangle \approx \sqrt{\frac{4 \pi N_{cells}}{N_{int}^{2}}} (1 + \frac{1}{2}\epsilon - \frac{1}{8}\epsilon^2).
\label{eq_mean_interface_edge_length_epsilon_expansion}
\end{equation}{}

Similarly, substituting $\langle a \rangle = 1 + \epsilon$ to Eq. \ref{eq_supp_lspring_01} and expanding yields
\begin{equation}
\begin{split}
    \langle l_{sp} \rangle &= \sqrt{\frac{2}{\sqrt{3}}} \sqrt{\frac{A_{box} - N_{cells}}{N_{sp}}} \sqrt{1 - \frac{N_{cells}}{A_{box} - N_{cells}} \epsilon} \\
    & \approx \sqrt{\frac{2}{\sqrt{3}}} b_1 ( 1 - \frac{1}{2} b_2 \epsilon - \frac{1}{8} b_2^2 \epsilon^2),
\label{eq_supp_lspring_02}
\end{split}{}
\end{equation}
where $b_1 = \sqrt{(A_{box} - N_{cells})/N_{sp}}$ and $b_2 = N_{cells}/(A_{box} - N_{cells})$. The total energy is now
\begin{multline}{}
    e_{tot} = N_{cells}\epsilon^{2} 
    \\+ \gamma \sqrt{4 \pi N_{cells}} (1 + \frac{1}{2} \epsilon - \frac{1}{8} \epsilon^{2})
    \\+ \frac{1}{2} k_{sp} N_{sp} \Bigg( \sqrt{\frac{2}{\sqrt{3}}} b_1 ( 1 - \frac{1}{2} b_2 \epsilon - \frac{1}{8} b_2^2 \epsilon^2) - l_{0} \Bigg) ^2
    \\+ k_{p} N_{cells} (\langle p \rangle - p_{0})^{2} .
\label{eq_nondim_energy_full_circular_epsilon_01}
\end{multline}

We assume that in this regime, $\langle p \rangle$ does not depend on $\langle l_{int} \rangle$ or $\langle l_{sp} \rangle$. The cell perimeters are not coupled to the total tissue perimeter in the way that cell areas are coupled to the total tissue area. Instead, for a given fixed cell area, the cell perimeter can vary, given the constraints on cell shape.

At this point, we simplify things by first taking the limit of no ECM tension, or, equivalently, $k_{sp} \rightarrow 0$. Combining powers of $\epsilon$ and completing the square then gives
\begin{multline}
    e_{tot} = (N_{cells} - \frac{1}{4} \gamma \sqrt{\pi N_{cells}}) \Bigg ( \epsilon + \frac{1}{\frac{\sqrt{4 \pi N_{cells}}}{ \pi} \frac{1}{\gamma} - \frac{1}{2}} \Bigg) ^{2} 
    \\ + k_{p} N_{cells} (\langle p \rangle - p_{0})^{2} + B,
\end{multline}
where $B$ is a collection of constants that simply shifts the total energy. Dropping this term, and putting the energy in terms of $\langle a \rangle$ again, we get
\begin{multline}
    e_{tot} = (N_{cells} - \frac{1}{4} \gamma \sqrt{\pi N_{cells}}) \Bigg ( \langle a \rangle - 1 + \frac{1}{\frac{\sqrt{4 \pi N_{cells}}}{ \pi} \frac{1}{\gamma} - \frac{1}{2}} \Bigg) ^{2} 
    \\ + k_{p} N_{cells} (\langle p \rangle - p_{0})^{2},
\end{multline}
which we rewrite as
\begin{equation}
    e_{tot} = k_{a}^\prime (\langle a \rangle - a_{0}^\prime)^{2} + k_{p} N_{cells} (\langle p \rangle - p_{0})^{2}
\end{equation}
and define 
\begin{equation}
    k_{a}^\prime = N_{cells} - \frac{1}{4} \gamma \sqrt{\pi N_{cells}}
\label{eq_kAprime}
\end{equation} 
and 
\begin{equation}
    a_{0}^\prime = 1 - \frac{1}{\frac{\sqrt{4 \pi N_{cells}}}{ \pi} \frac{1}{\gamma} - \frac{1}{2}}.
\label{eq_a0prime}
\end{equation}

We now divide both sides by a (dimensionless) constant, $k_{A}^\prime a_{0}^{\prime 2}$ to arrive at
\begin{equation}
    \tilde{e}_{tot} = (\tilde{\langle a \rangle} - 1)^{2} + \tilde{k}_{p} (\tilde{\langle p \rangle} - \tilde{p}_{0})^{2},
\label{eq_supp_reformulated_energy_func}
\end{equation}
which is the same as Eq. 2 in the main text. The effective model parameters are denoted by tildes and defined in Table \ref{tbl_params_02}. They are each functions of $k_{a}^{\prime}$ and/or $a_{0}^{\prime}$, which are defined in Eqs. \ref{eq_kAprime} and \ref{eq_a0prime}. It is this effective model that we use to produce the results in Fig. 2 (c) and (d) of the main text.

\begin{table}{}
\begin{center}
\renewcommand{\arraystretch}{1.5} 
\begin{tabular}{c|c}
  Effective & Original \\
  \hline
  \hline
    $\tilde{e}_{total}$ & $\frac{e_{total}}{k_{a}^{\prime}a_{0}^{\prime 2}}$ \\
    $\tilde{\langle a \rangle}$ & $\frac{\langle a \rangle}{a_{0}^{\prime}}$ \\
    $\tilde{\langle p \rangle}$ & $\frac{\langle p \rangle}{\sqrt{a_{0}^{\prime}}}$ \\
    $\tilde{p}_{0}$ & $\frac{p_{0}}{\sqrt{a_{0}^{\prime}}}$ \\
    $\tilde{k}_{p}$ & $\frac{k_{p} N_{cells}}{k_{a}^{\prime}a_{0}^{\prime}}$ \\
\end{tabular}{}
\caption{\label{tbl_params_02} The mapping from the new, effective model parameters to the original, dimensionless model parameters.}
\end{center}{}
\end{table}

\begin{figure}
    \centering
    \includegraphics[width=0.4\textwidth]{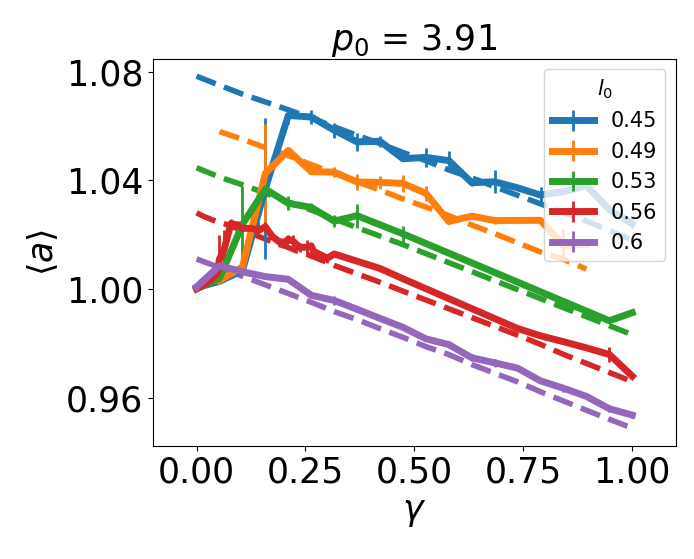}
    \caption{The mean cell area, $\langle a \rangle$, as a function of interfacial tension, $\gamma$, for varying spring rest length, $l_{0}$. Simulation results are shown in solid lines and the theory is shown in dashed lines.}
    \label{fig_supp_full_circular_regime_model}
\end{figure}

If we do not ignore the third term in Eq. \ref{eq_nondim_energy_full_circular_epsilon_01}, we can expand it, drop terms greater than $O(\epsilon^2)$, again collect terms that are linear and quadratic in $\epsilon$, complete the square, and divide by a new $a_{0}^\prime$. This yields Eq. \ref{eq_supp_reformulated_energy_func}, but with new definitions of $\tilde{p}_{0}$ and $\tilde{k}_{p}$. We do not detail the algebra here, but show the results of this reformulation in Fig. \ref{fig_supp_full_circular_regime_model}, which demonstrates the behavior of the mean cell area, $\langle a \rangle$, as a function of $\gamma$, for a set of $l_{0}$ values and fixed $p_{0}=3.91$. The measured values from simulations are shown as solid curves. We assume that the cells reach their preferred areas ($\langle a \rangle = a_{0}^\prime$) and therefore plot $a_{0}^\prime(\gamma, l_{0}, k_{sp}, N_{cells}, N_{springs}, A_{box})$ from our model, on top of the observed values, using dashed curves.
{}
Although our analytical model ignores the disorder and fluctuations in our simulations, we find good agreement between the predicted and measured values for the cell areas.

\subsection{Analytical model: predicting the critical interfacial tension}
\label{sec_prediction_gamma_c_mf}

In Fig. \ref{fig_supp_full_circular_regime_model}, it is clear that the model described in Sec. \ref{sec_regime_1_math_details}, represented by dashed lines, breaks down below a specific value of $\gamma$ for each curve. We call this special point $\gamma_{c}$, and define it as the value corresponding to the peak of the mean cell area, $\langle a \rangle$, vs. $\gamma$ curve, for a fixed $p_{0}$ and $l_{0}$. For $\gamma > \gamma_{c}$, the tissue is compact, and \textit{compact tissue regime} model works well to describe $\langle a \rangle$. For $\gamma < \gamma_{c}$, the tissue is irregular and cavities are present, and specifically for $\gamma \approx 0$, our \textit{irregular tissue regime} model works well to describe the cell shapes (see Sec. \ref{sec_supp_regime_2_analysis}). At $\gamma_c$, the system clearly switches sharply from one of these regimes to the other, and we hypothesize that this sharpness is due to a sudden change in whether cavities are energetically-favorable. In other words, $\gamma_c$ signifies the onset of an instability in the tissue-ECM interface.

To test this hypothesis, we devise a simple toy model, in which the tissue is circular, with radius, $R$, and contains a cavity, also circular, of radius, $r$. The energetic contribution from the boundary is
\begin{equation}
    E_{int} = \gamma (2 \pi r + 2 \pi R).
\label{eq_supp_gamma_c_E_int}
\end{equation}{}
We approximate the energetic contribution from the spring network as
\begin{equation}
    E_{sp.net} = \frac{1}{2} k_{sp} N_{sp} (\langle l_{sp} \rangle - l_{0})^2.
\label{eq_supp_gamma_c_E_spring_net}
\end{equation}{}
The total area of the spring network is related to $R$ as
\begin{equation}
    A_{sp.net} = A_{box} - \pi R^2,
\label{eq_supp_gamma_c_A_spring_net}
\end{equation}{}
and therefore, using the same description of the spring network as in Sec. \ref{sec_regime_1_math_details} and applying Eq.  \ref{eq_supp_area_spring_net_01}, the mean spring length, $\langle l_{sp} \rangle$, is
\begin{equation}
    \langle l_{sp} \rangle = \sqrt{\frac{2}{\sqrt{3}}} \sqrt{\frac{A_{box} - \pi R^2}{N_{sp}}}.
\label{eq_supp_gamma_c_lsp}
\end{equation}{}

We assume the energetic contributions from the tension in the spring network and the interfacial tension dominate, and therefore deviations in cell areas and perimeters are small. Specifically, we let $\langle a \rangle = 1$, and therefore
\begin{equation}
    A_{tissue} = \pi R^2 - \pi r^2 = N_{cells},
\label{eq_supp_gamma_c_Atissue}
\end{equation}{}
which allows us to write $R$ as a function of $r$:
\begin{equation}
    R = \sqrt{\frac{N_{cells}}{\pi} + r^2}.
\label{eq_supp_gamma_c_R}
\end{equation}{}
The total energy is now 
\begin{multline}
    E_{tot}(r, \gamma, l_{0}) = 2 \pi \gamma (r + R(r)) 
    \\+ \frac{1}{2} k_{sp} N_{sp} (\langle l_{sp} \rangle(r) - l_{0})^2.
\label{eq_supp_gamma_c_E_tot}
\end{multline}{}

\begin{figure}
    \centering
    \includegraphics[width=0.4\textwidth]{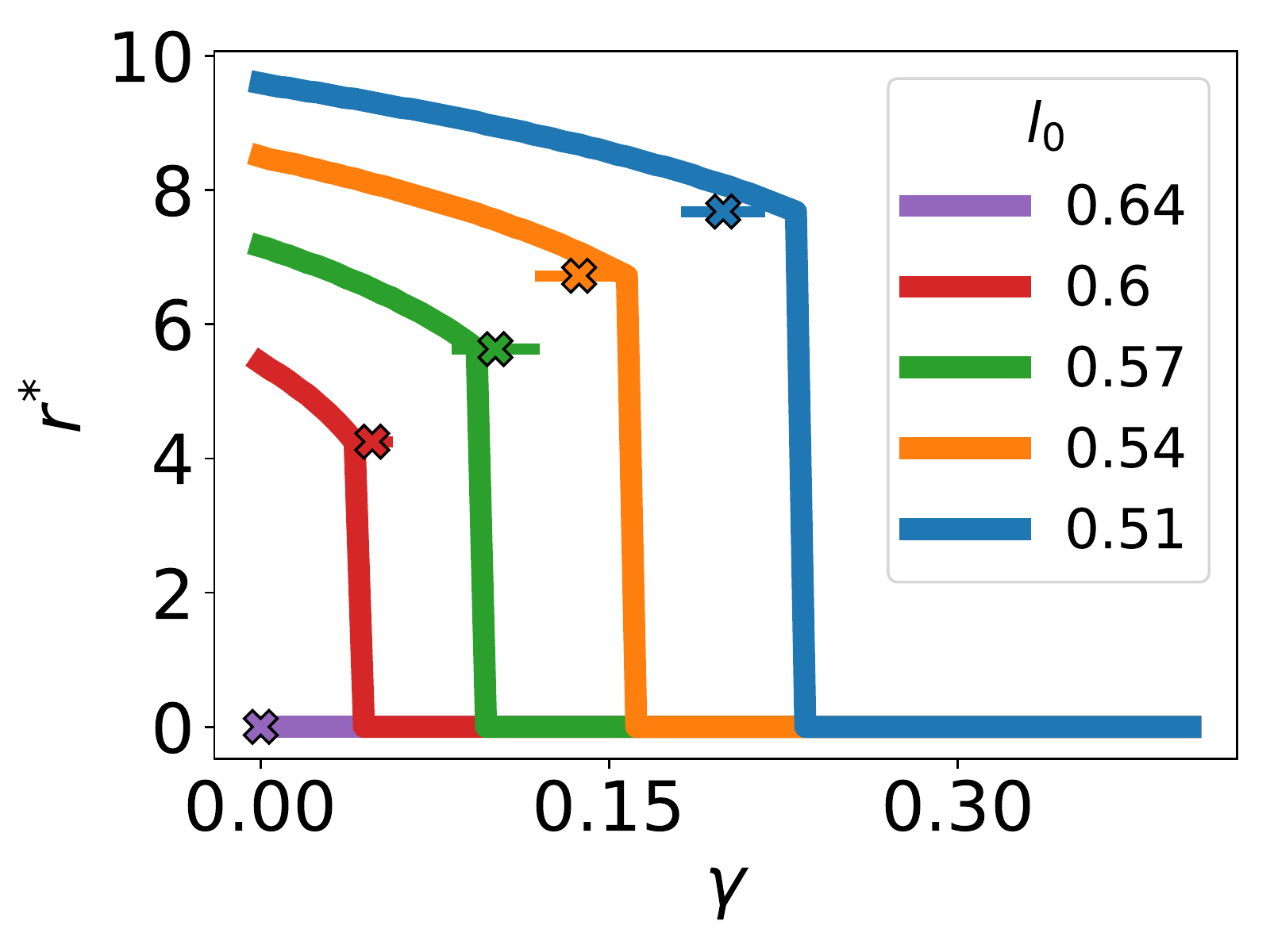}
    \caption{The instability model prediction for the critical cavity radius, $r^*$, as a function of interfacial tension, $\gamma$, is shown in solid lines for varying spring rest length, $l_{0}$, values. Each X indicates the critical $\gamma$ value, $\gamma_c$, extracted from the maximum of the $\langle a \rangle$ vs. $\gamma$ curve for the corresponding $l_{0}$. $\gamma_c$ values and their error bars are found by binning data points near each maximum and taking the largest bin mean and its standard deviation.}
    \label{fig_supp_rstar_vs_gamma}
\end{figure}{}

We minimize this energy with respect to $r$, for fixed $l_{0}$ and $\gamma$, and solve for $r^*$-- the value of the cavity radius, $r$, that minimizes the energy. The results are shown as the solid curves in Fig. \ref{fig_supp_rstar_vs_gamma}. For each fixed $l_{0}$ value, we find and plot $r^*$ for a range of $\gamma$ values. 

For each $l_{0}$ value, there is a jump in $r^*$ at a particular value of $\gamma$. For $\gamma$ above this point, the energetically-favorable cavity radius is zero, and the system prefers circular, compact tissue geometries. Immediately below this point, however, the system suddenly prefers a cavity of finite radius, which continues to grow as $\gamma$ decreases further. To compare the $\gamma$ value at which we see a jump to $\gamma_c$, we also plot the location of the peaks of $\langle a \rangle$ vs. $\gamma$ curves as Xs. We find good agreement between the $\gamma$ values associated with the peaks and the points at which $r^*$ becomes non-zero, meaning that our instability hypothesis successfully predicts $\gamma_c$.

\subsection{Analytical model: irregular tissue regime}
\label{sec_supp_regime_2_analysis}

\begin{figure}
    \centering
    \includegraphics[width=0.45\textwidth]{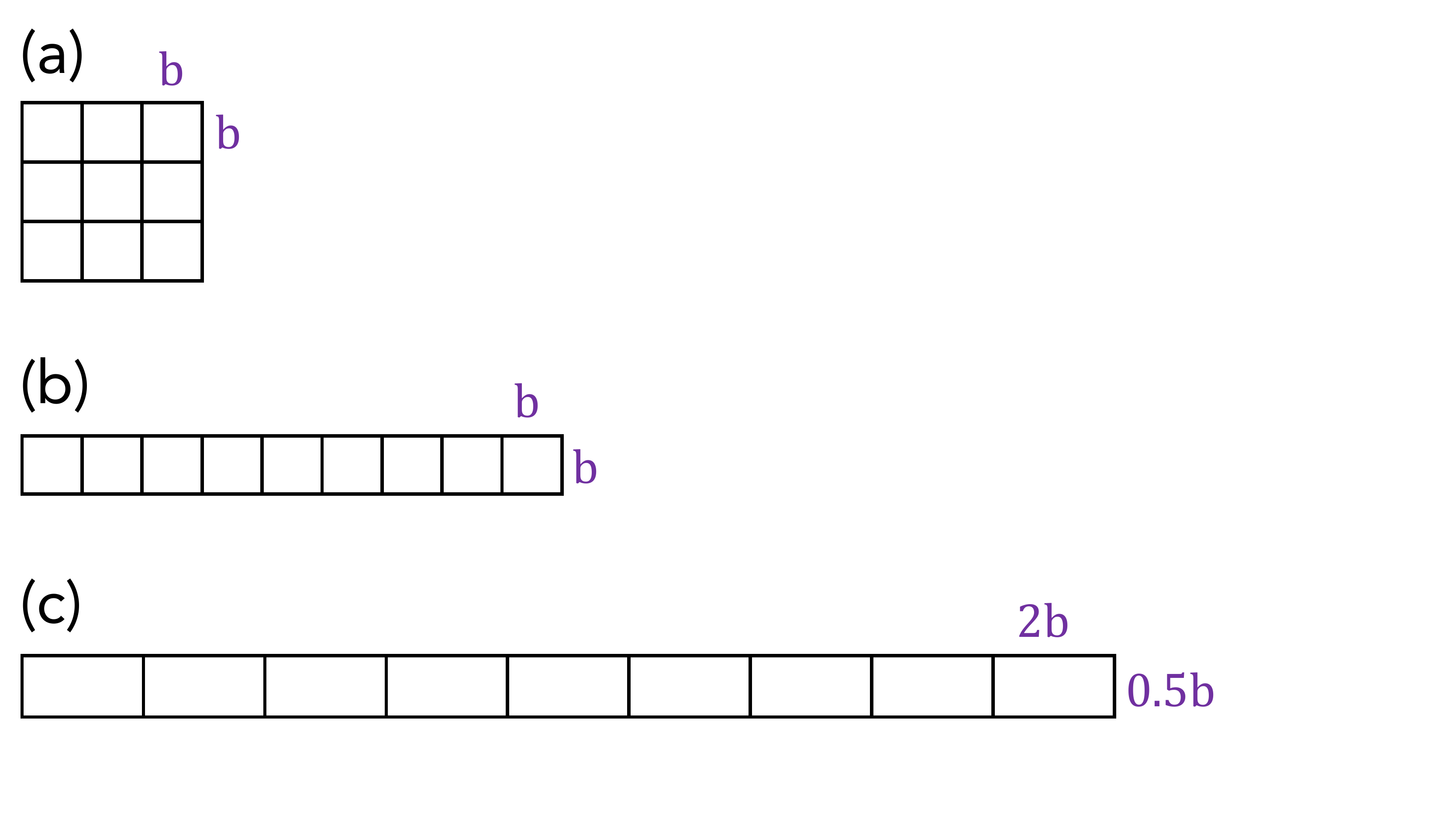}
    \caption{Illustration of toy model used to explain the increase in mean cell shape in the \textit{irregular tissue regime}.}
    \label{fig_regime_2_supp_illustration}
\end{figure}

When the surrounding spring network tension is high, the minimum energy configurations for the tissue become extremely irregular. We observe an increase in the perimeter of the cells with increasing spring network tension and with increasing cavity area between the spring network and tissue. To understand this, we employ yet another toy model. We start with a collection of square bricks, with side length $b$, compactly arranged as in Fig. \ref{fig_regime_2_supp_illustration}, (a). In our tissue-spring network simulations, the system is driven to decrease the spring network area and does this by creating more tissue-ECM boundary. In our toy model system, we can increase the total boundary by simply arranging the blocks in a line, as in Fig. \ref{fig_regime_2_supp_illustration}, (b). If we now want to increase the boundary even more, we must start to deform the blocks. To do this without changing the area of each individual block (which would incur an energetic cost in the actual system), we simply expand each block along the line and shrink it perpendicular to the line in such a way that the area does not change, as shown in Fig. \ref{fig_regime_2_supp_illustration}, (c). As the sides of the blocks perpendicular to the line shrink and the parallel sides grow, the perimeter of each block becomes proportional to the length of the line.

{}

If we now imagine wrapping this line around a cavity, its length becomes the circumference of this cavity, and each block's perimeter is therefore proportional to this circumference, and in turn, the radius, of the cavity. This means that each block's perimeter is proportional to the square root of the area of the cavity. Multiplying this cavity area by some value gives us the total cavity area, $A_{cav}$, and we can write
\begin{equation}
   \langle p \rangle = B_{1} \sqrt{A_{cav}} + B_{2}.
\label{eq_supp_case_02_cell_perim_final}
\end{equation}
Since the shape, $s$, is defined as $s=p/\sqrt{a}$, and we assume $a=a_{0}=1$, we can replace $\langle p \rangle$ with $\langle s \rangle$ and we arrive at Eq. 3 of the main text.

\subsection{Defining and computing the cell-cell alignment parameter}
\label{sec_supp_Q_parameter}

To quantify the cell-cell alignment, we define and compute $Q$, as in Ref. \cite{Wang_2019}. The value of this quantity indicates how aligned a set of cells are, modulated by their circularity. In other words, unlike a simple nematic order parameter, $Q$ weighs the alignment of highly polarized shapes more than that of less polarized shapes, even if the orientations of their axes are identical. This assures that regular, symmetric shapes are assigned zero alignment, regardless of their orientation relative to each other.

To compute $Q$ for a set of cells, one first triangulates the system with each vertex shared by three cells. By connecting the centers of these three cells, we produce a triangle. Doing this for each vertex in our system produces a set of triangles-- the dual of the original network of vertices. The general idea is now to determine the deformation of each triangle from some reference triangle, and compare these deformations to see how similarly-elongated the triangles in the set are. The details are as follows:

\begin{enumerate}

    \item Define a reference triangle. For example, an equilateral triangle with area 1. Its side length is given by
    \begin{equation*}
        d_{ref} = \frac{2}{\sqrt{\sqrt{3}}}
    \end{equation*}{}
    and a matrix that defines it is given by
    \begin{equation*}
    m_{2} = d_{ref} \times \frac{1}{2} \times
        \begin{bmatrix}
        -\sqrt{3} & -\sqrt{3}\\
        1 & -1
    \end{bmatrix}
    \end{equation*}{}
    \item For each triangle, list the coordinates of its corners as $[r_{A}, r_{B}, r_{C}]$, such that they are in counter-clockwise order.
    \item Construct a matrix that defines the current triangle:
    \begin{equation*}
    m_{2} =
        \begin{bmatrix}
        r_{x, B} - r_{x, A} & r_{x, C} - r_{x, A}\\
        r_{y, B} - r_{y, A} & r_{y, C} - r_{y, A}
    \end{bmatrix}
    \end{equation*}{}
    \item Compute the shape tensor, $S$, of the current triangle, which is defined as
    \begin{equation*}
    S = m_{1} \times m_{2}
    \end{equation*}{}
    \item Compute the trace part, $t$, of $S$:
    \begin{equation*}
    t = \frac{1}{2}\text{Tr}[S] \times
    \begin{bmatrix}
        1 & 0\\
        0 & 1
    \end{bmatrix}
    \end{equation*}{}
    \item Compute the symmetric, traceless part of $S$:
    \begin{equation*}
    S_{symm} = \frac{1}{2}(S + S^{\text{T}}) - t
    \end{equation*}{}
    \item Compute the antisymmetric part of $S$ (this is already traceless):
    \begin{equation*}
    S_{asymm} = \frac{1}{2}(S - S^{\text{T}})
    \end{equation*}{}
    \item Compute the rotation angle, $\theta$, of the triangle. In programming languages with the arctan2 function, this can be found as 
    \begin{equation*}
    \theta = \text{arctan2}(S_{asymm, yx}, t_{xx})
    \end{equation*}{}
    \item Compute the ``norm'' of the symmetric, traceless part of $S$. Note this is not defined in the standard way.
    \begin{equation*}
    |S_{symm}| = (S^{2}_{symm, xx} + S^{2}_{symm, xy})^{1/2}
    \end{equation*}{}
    \item Compute the determinant of $S$, $\text{det}(S)$.
    \item Finally, compute the elongation tensor of the triangle, $q$:
    \begin{equation*}
        q = \frac{1}{|S_{symm}|} \sinh^{-1}{(\frac{|S_{symm}|}{\sqrt{\text{det}(S)}})}(S_{symm} \times R)
    \end{equation*}{}
    where $R$ is the 2D, clockwise rotation matrix:
    \begin{equation*}
        R = \begin{bmatrix}
        \cos{\theta} & \sin{\theta}\\
        -\sin{\theta} & \cos{\theta}
    \end{bmatrix}
    \end{equation*}{}
    \item $Q$ of the system is defined as the weighted average of all $q$ tensors (for all triangles):
    \begin{equation*}
        Q = \frac{\sum_{i} a_{i} q_{i}}{\sum_{i} a_{i}}
    \end{equation*}{}
    where the sum is over triangles, $i$, and $a_{i}$ is the area of triangle $i$. The area of triangle with its corners listed in counter-clockwise order can be computed using the ``shoelace formula'' as:
    \begin{multline*}
        a_{i} = \frac{1}{2}(r_{x, A}(r_{y, B} - r_{y, C}) \\
        + r_{x, B}(r_{y, C} - r_{y, A})\\
        + r_{x, C}(r_{y, A} - r_{y, B}))
    \end{multline*}
    \item Q is a tensor, and we report the norm of it, given by
    \begin{equation*}
        |Q| = (Q^{2}_{xx} + Q^{2}_{yy})^{1/2}
    \end{equation*}{}

\end{enumerate}

This procedure can be done for any group of cells (and corresponding group of triangles). In the \textit{irregular tissue regime}, our tissue is being expanded roughly radially by the tension in the surrounding network. The cells are not all being deformed in the same direction and we therefore do not expect the system to be globally aligned. Instead, high cell-cell alignment in our system means high \textit{local} alignment. To measure this, we define, for each cell, a set of triangles using the centers of the cell and its neighbors. We compute $Q$ for this set and then average over all sets. This gives a measure of the mean local alignment for a given tissue configuration.

\subsection{Measuring T1 energy barriers}
\label{sec_supp_t1_barriers}

In addition to computing the effective diffusivity, $D_{eff}$, of the cells at finite temperature, the phase of the tissue can be measured at zero temperature, by analyzing its response to a non-linear deformation, such as a T1 transformation. In other words, if there is a non-zero energy barrier to undergoing a T1, the system is solid-like \cite{Bi_2014, Sahu_2019}. Although the trends of increasing shape and alignment with increasing fluidity remain intact, we find that in the \textit{irregular tissue regime}, the quantitative measures of mean cell shape and cell-cell alignment are sensitive to temperature. Moreover, $D_{eff}$ is sensitive to finite-size effects, while the calculation of T1 energy barriers is not (see Sec. \ref{sec_supp_finite_size_all}). Therefore, in an attempt to decouple the effects of temperature and system size from tissue fluidity, we compute the energy barriers for cells to undergo T1s at zero temperature, and compare the results to our finite-temperature measure of fluidity, $D_{eff}$.

To measure a T1 barrier, we minimize the total energy of the system using the FIRE algorithm (see sec. \ref{sec_energy_minimization_methods}), choose a cell edge at random, and then incrementally shorten the edge, minimizing the system energy again after each shortening step, and stop when the edge length equals 0.006. The difference between the final and initial energies is taken to be an estimate of the barrier height to executing that T1. Averaging over many edges gives us a mean barrier height for a system with given parameter values, such as $p_{0}$ of the tissue and $l_{0}$ of the surrounding spring network.

\begin{figure}
    \centering
    \includegraphics[width=0.4\textwidth]{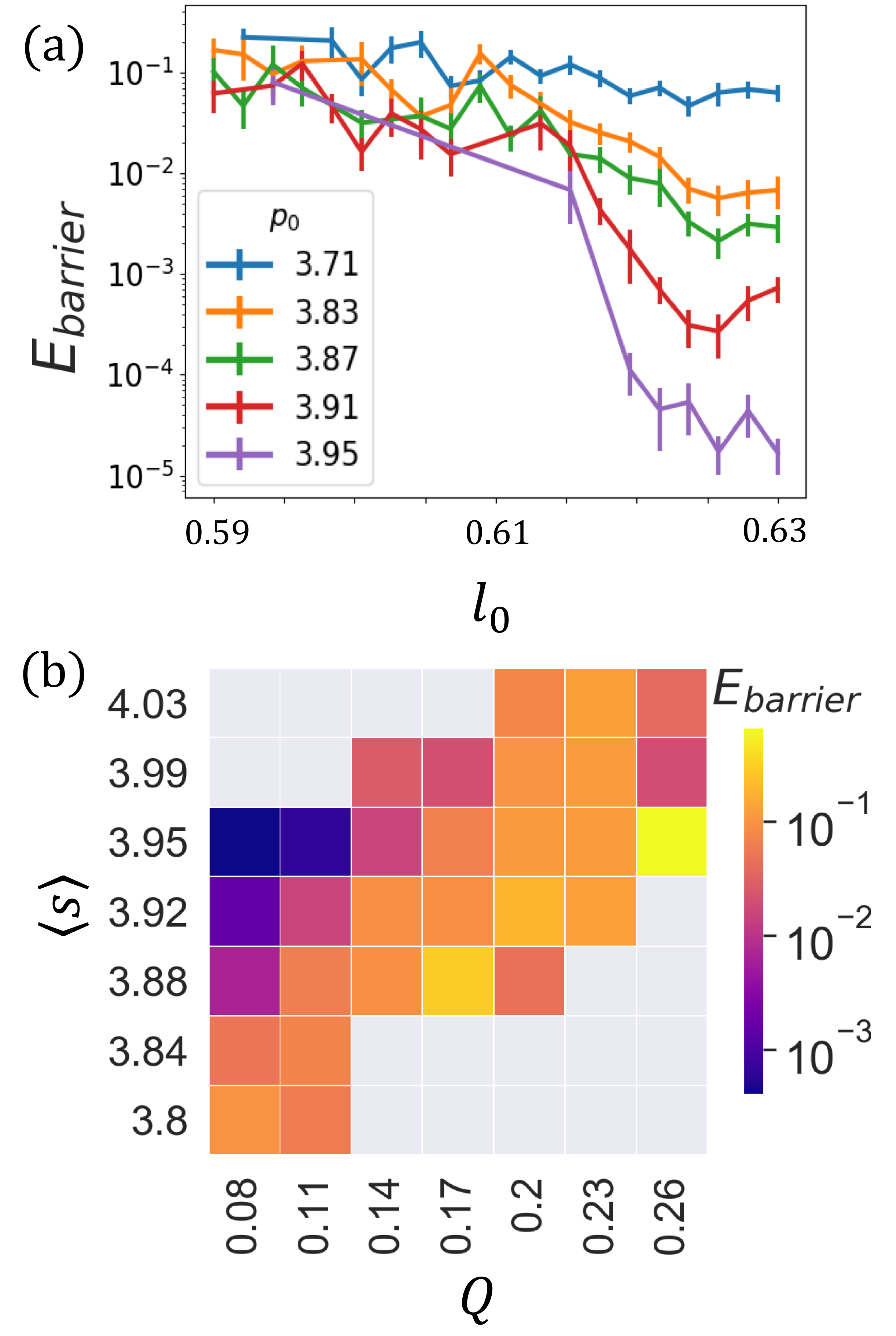}
    \caption{T1 energy barrier as a function of spring rest length, mean cell shape, and cell-cell alignment. (a) The energy barrier, $E_{barrier}$, for a tissue cell edge to undergo a T1 transition, as a function of the rest length, $l_{0}$, of the springs in the ECM. For $l_{0} > 0.62$, the ECM is floppy, and $E_{barrier}$ decreases from approximately $1\times10^{-1}$ for $p_{0}=3.71$ to $1\times10^{-5}$ for $p_{0}=3.95$. For approximately $l_{0} < 0.62$, the ECM is rigid, and $E_{barrier}$ approaches $1\times10^{-1}$ for all $p_{0}$ values. (b) A heatmap of $E_{barrier}$ as a function of the measured mean shape, $\langle s \rangle$, and cell-cell alignment, $Q$, of the tissue cells. This diagram is made by binning the results of simulations across ranges of $p_{0}$ and $l_{0}$.}
    \label{fig_supp_ebarriers}
\end{figure}

We find that for a floppy ECM, the energy barrier is approximately $1 \times 10^{-5}$ for a tissue with $p_{0}=3.95$, and increases as $p_{0}$ increases, to approximately $1 \times 10^{-1}$ for $p_{0}=3.71$, as seen previously for bulk tissue \cite{Sahu_2019}. As the ECM stiffens, however, the energy barriers for all $p_{0}$ values approach $1 \times 10^{-1}$, indicating a solidification of the tissue. This is shown in Fig. \ref{fig_supp_ebarriers} (a). For each $l_{0}$-$p_{0}$ pair, we also measure the mean cell shape and cell-cell alignment parameter, $Q$, and find that high cell shapes and low alignment correspond to low energy barriers (fluid-like tissue), whereas high cell shapes and high alignment correspond to higher barriers (more solid-like tissue), as shown in Fig. \ref{fig_supp_ebarriers} (b). These results corroborate our findings for $D_{eff}$ reported in the main text.

{}

\subsection{Finite-size effects}
\label{sec_supp_finite_size_all}

\begin{figure}
    \centering
    \includegraphics[width=0.4\textwidth]{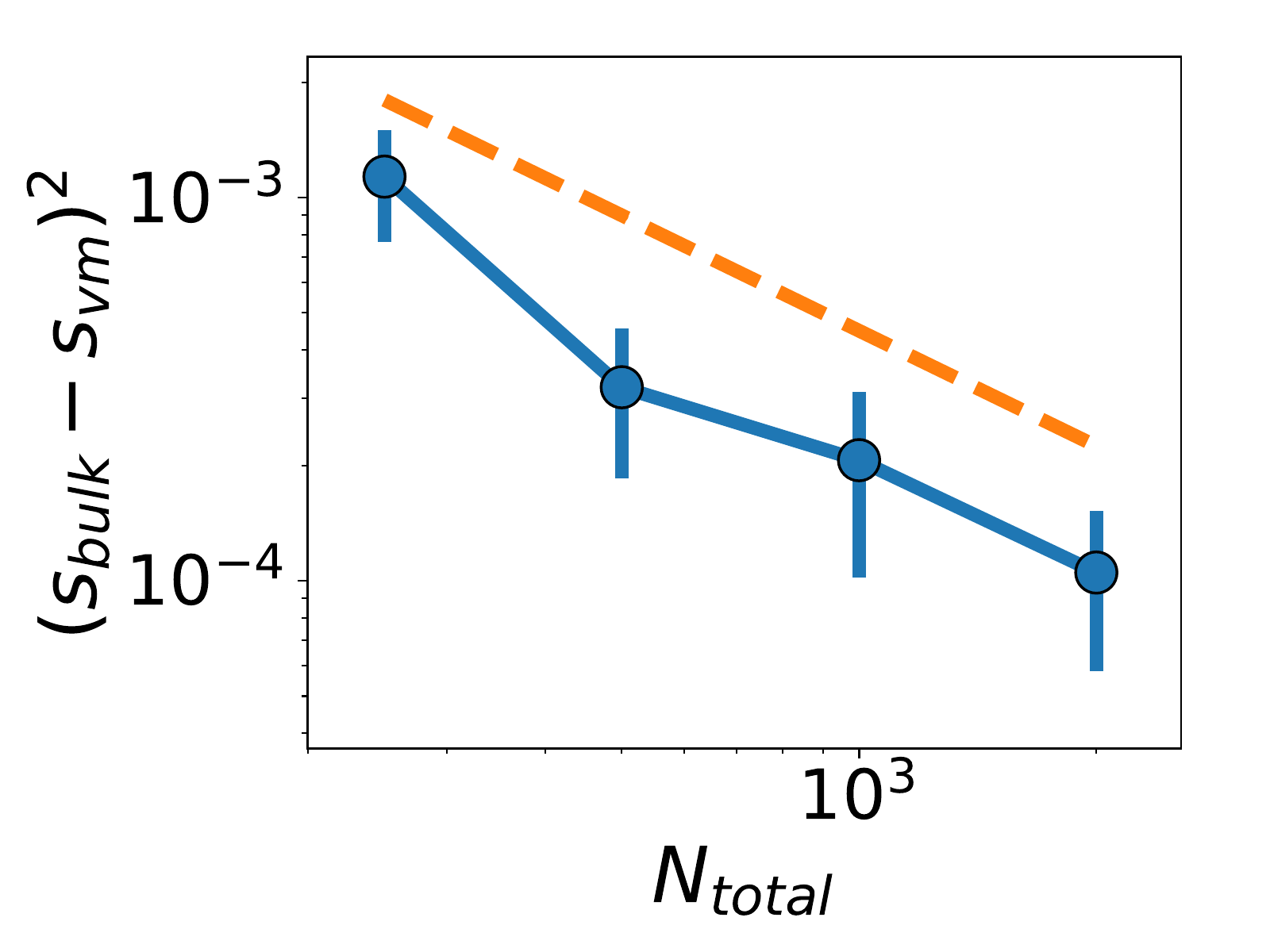}
    \caption{The squared difference of the mean cell shapes, between spheroid and all-cell simulations, for varying system sizes. In each spheroid simulation, the tissue takes up 20$\%$ of the total system size. The dashed line has a slope of -1 on this log-log plot, illustrating the roughly $N^{-1}$ behavior.}
    \label{fig_supp_l2norm_regime_01}
\end{figure}

To understand the system-size dependence of our results, we first compute the mean squared difference between the average cell shapes for our spheroid embedded in ECM and the average cell shapes of a pure vertex model, for floppy network and varying interfacial tension, $\gamma$, as a function of total system size. In other words, we study how the mean squared difference between the red and blue curves in Fig. 2 (d) of the main text depends on system size, for fixed ratio of tissue area to box area of 20\%. The results are shown in Fig. \ref{fig_supp_l2norm_regime_01}. We find that the difference between the mean cell shape of a spheroid and that of an all-cell vertex model decreases as approximately $1/N_{system}$.

{}

\begin{figure}
    \centering
    \includegraphics[width=0.4\textwidth]{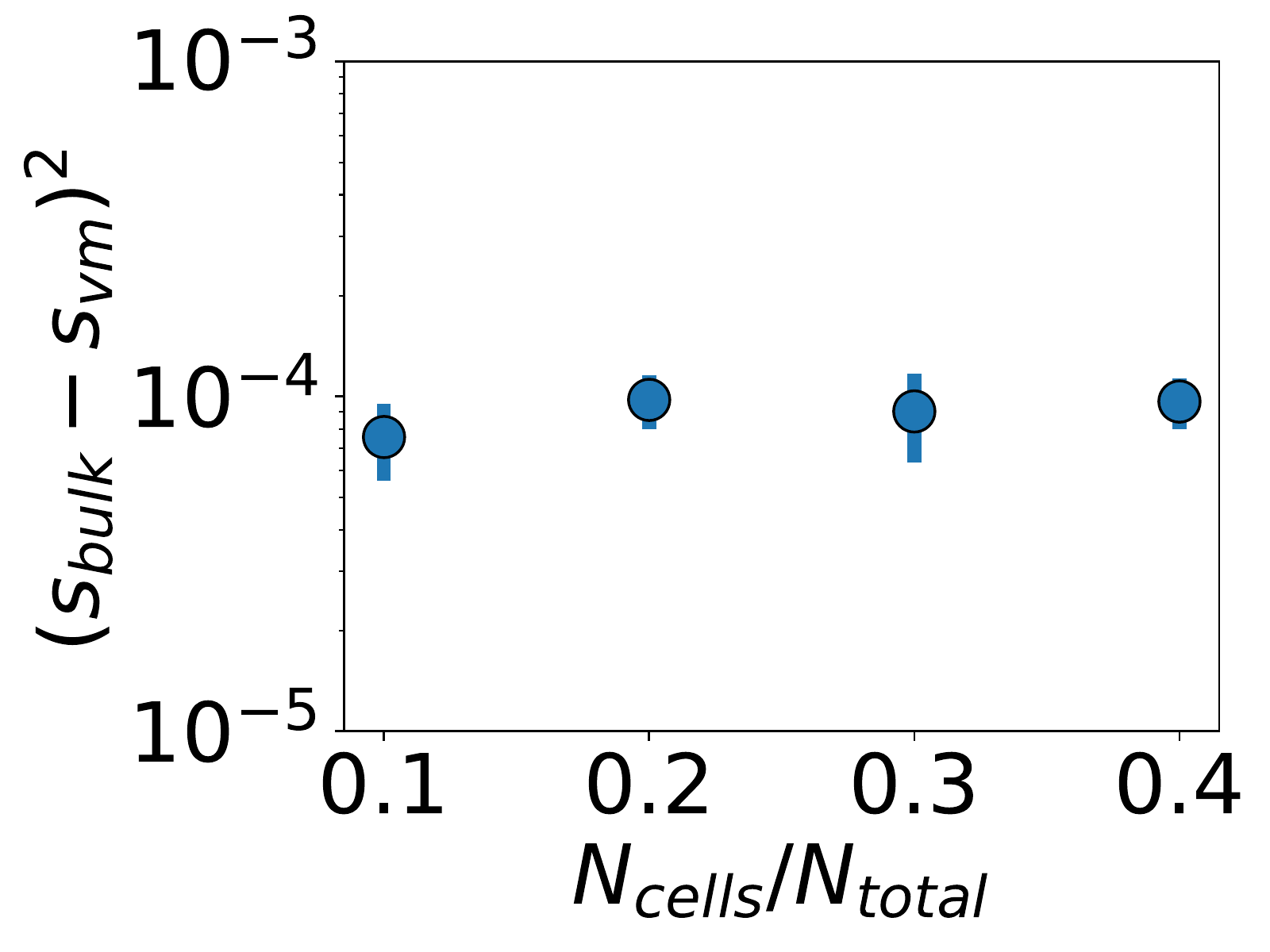}
    \caption{The squared difference of the mean cell shapes, between spheroid and all-cell simulations, for varying box sizes, for fixed $N_{cells}=800$.}
    \label{fig_supp_l2norm_ratio_regime_01}
\end{figure}

Throughout our work, in order to reduce the number of free parameters in our model, we've maintained a constant ratio of $N_{cells}/N_{total}=0.2$, where $N_{total}$ is the total number of polygons in the simulation box. However, we expect our results to depend on this choice. To understand the extent of this effect, we repeat the above analysis, now for fixed $N_{cells}=800$ and varying ratios of $N_{cells}/N_{total}$. As shown in Fig. \ref{fig_supp_l2norm_ratio_regime_01}, the squared difference of our model from the bulk vertex model does not in fact depend significantly on the ratio of tissue size to box size. For a fixed tissue size of $N_{cells}=800$, the tissue cell behavior is similar to the bulk vertex model, regardless of the relative size of the box.

{}

\begin{figure}
    \centering
    \includegraphics[width=0.4\textwidth]{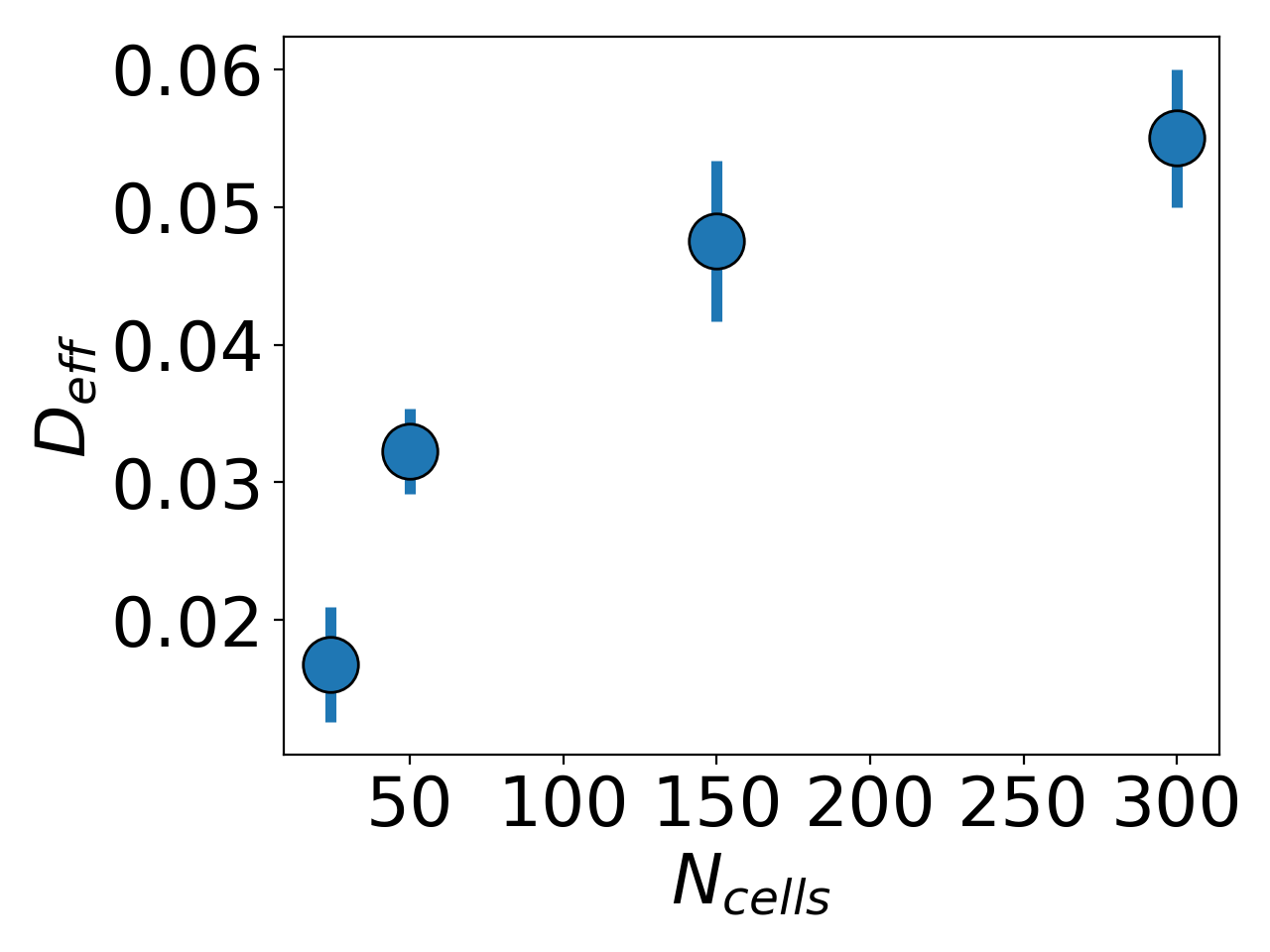}
    \caption{Effective diffusion coefficient, $D_{eff}$, vs. system size for a tissue of $p_{0}=3.9$ and floppy surrounding network. As the system size decreases, $D_{eff}$ decreases, although all other system parameters are fixed.}
    \label{fig_supp_deff_system_size_floppy_network_regime_02}
\end{figure}

We also find that $D_{eff}$ depends on system size. With a finite boundary, $D_{eff}$ decreases as the system size is decreased, when surrounded by floppy network, as show in Fig. \ref{fig_supp_deff_system_size_floppy_network_regime_02}.

{}

\begin{figure}
    \centering
    \includegraphics[width=0.4\textwidth]{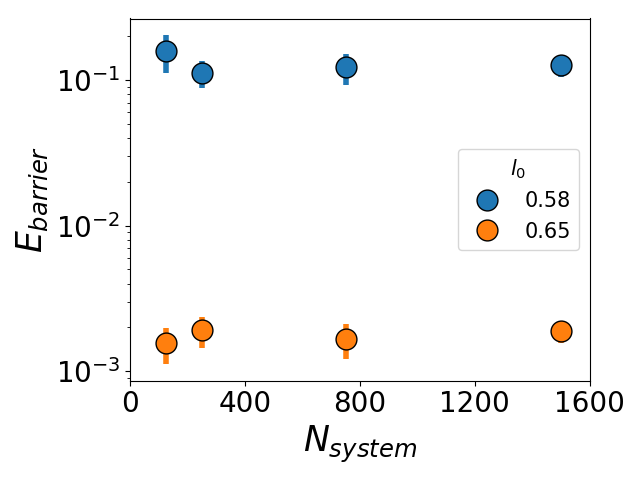}
    \caption{The energy barrier, $E_{barrier}$, for a cell edge undergoing a T1 vs. system size for a tissue of $p_{0}=3.9$, for both floppy ($l_{0}=0.65$) and rigid ($l_{0}=0.58$) surrounding ECM. For a rigid network, $E_{barrier}$ is about two orders of magnitude greater than for a floppy network, regardless of the system size.}
    \label{fig_supp_Eb_system_size_regime_02}
\end{figure}

As we increase the network tension, cavities are formed and the tissue spreads into ``channels'' that surround these cavities. To avoid conflating the finite-size effect on $D_{eff}$ with the potential rigidifying effect of the surrounding network tension, we also quantify the phase of the tissue using T1 energy barriers (see Sec. \ref{sec_supp_t1_barriers}), which do not suffer from finite-size effects, as show in Fig. \ref{fig_supp_Eb_system_size_regime_02}. We find that as $l_{0}$ decreases and the tension in the network increases, the energy barriers increase, supporting the argument that the simultaneous decrease in $D_{eff}$ is due to the interaction with the rigid ECM and not a finite-size effect.

{}

\subsection{Temperature dependence}
\label{sec_temperature_dependence}

Increasing the simulation temperature, $T$, increases the magnitude of fluctuations, adding forces of increasing strength and random direction to each vertex (see Sec. \ref{sec_brownian_dynamics_methods}). In the bulk vertex model, adding thermal energy can enable cells to overcome T1 energy barriers, even for $p_{0}<3.81$, and induce high cell shapes, as thermal forces perturb the positions of the cell vertices. In our model, we expect to see a similar trend in the cell shape as a function of temperature. However, these thermal forces are now also competing with the forces due to interfacial tension and ECM tension. 
\begin{figure}
    \centering
    \includegraphics[width=0.4\textwidth]{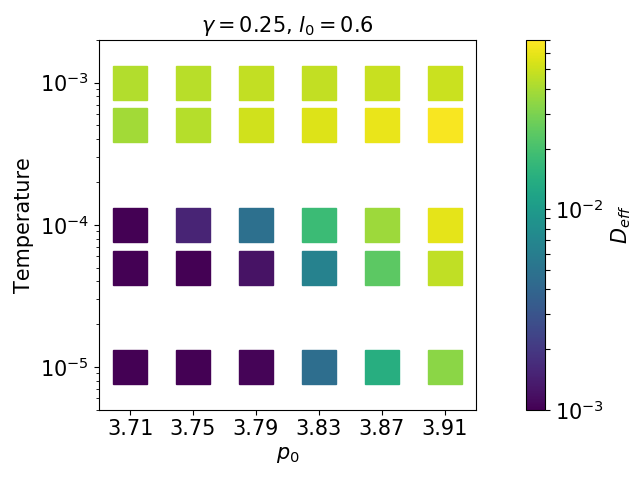}
    \caption{Effective diffusion coefficient, $D_{eff}$, as a function of simulation temperature and $p_{0}$, for a ``balanced'' pair of $l_{0}$ and $\gamma$ values. When $\gamma$ and $l_{0}$ are balanced, temperature effects the phase of the tissue, as a function of $p_{0}$, as it does in the bulk vertex model.}
    \label{fig_supp_temp_vs_p0_vs_deff}
\end{figure}{}
In order to understand this competition, we first check that temperature is playing a fundamentally similar role in our model. We identify a pair of interfacial tension, $\gamma$, and equilibrium spring length, $l_{0}$, values for which the tissue is neither expanded or compressed. In other words, for a chosen $l_{0}$, we identify the $\gamma$ value for which the pressures generated by the surface and by the external network are balanced, meaning that the tissue is compact and the mean cell shape is what we would expect for bulk tissue for the current $p_{0}$. Fig. \ref{fig_supp_temp_vs_p0_vs_deff} demonstrates the dependence of the diffusivity, $D_{eff}$, on $T$ and $p_{0}$, for a balanced pair of $\gamma$ and $l_{0}$ of 0.25 and 0.6, respectively. The behavior at $T=1e-5$ confirms that the pressures on the tissue are balanced for this pair of $\gamma$ and $l_{0}$, as there we see a transition in $D_{eff}$ around $p_{0} \approx 3.81$, matching the bulk vertex model. We find that as we increase $T$, the transition point in $p_{0}$ decreases. This indicates that, like the bulk vertex model, thermal fluctuations effectively decrease the energy barriers for cell rearrangements, fluidizing the tissue ``sooner'' (at lower values of $p_{0}$). 

\begin{figure*}[ht]
    \centering
    \includegraphics[width=0.9\textwidth]{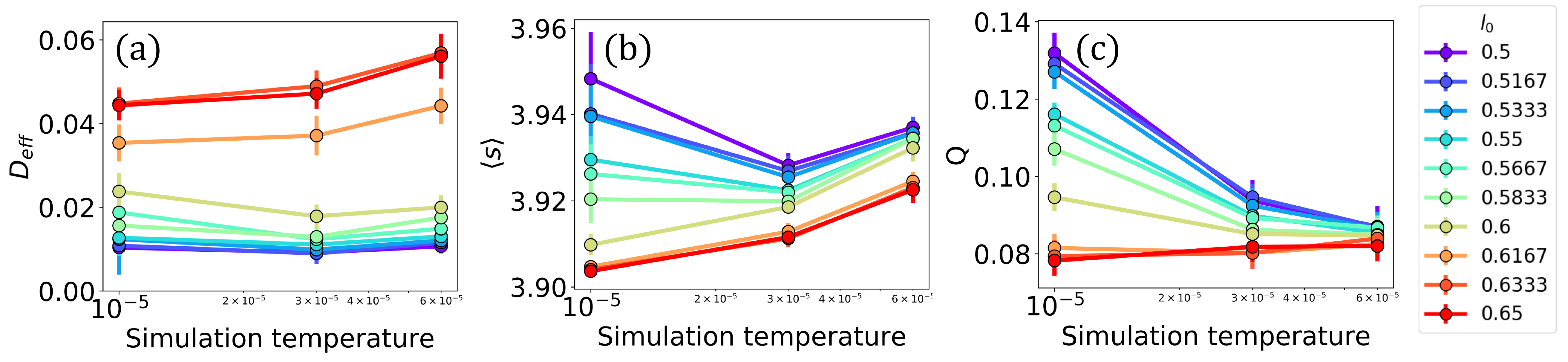}
    \caption{The temperature-dependence of diffusivity, $D_{eff}$, mean cell shape, $\langle s \rangle$, and cell-cell alignment, $Q$. Here, $\gamma=0$ and $l_{0}$ varies from values corresponding to rigid ECM (purple to yellow curves) to floppy network (orange to red curves). For a fixed temperature, increasing the rigidity (decreasing $l_{0}$) decreases $D_{eff}$ and increases both $\langle s \rangle$ and $Q$. The differences in $\langle s \rangle$ and $Q$ across $l_{0}$ values are less pronounced at higher temperatures, while the effect on $D_{eff}$ is relatively constant. While increasing thermal fluctuations mitigates the influence of external network tension on the tissue, it also promotes higher cell shapes and higher diffusion.}
    \label{fig_supp_temp_vs_deff_shape_q}
\end{figure*}{}

Having confirmed the role of temperature alone on the tissue, we now analyze the competition between increasing temperature and increasing ECM tension. At $T=0$, as $l_{0}$ decreases and ECM tension increases, we see the mean cell shape, $\langle s \rangle$, and cell-cell alignment, $Q$, both increase. Simultaneously increasing $T$ in this case has two competing effects: first, as in bulk tissue and when the ECM tension is balanced by interfacial tension, large fluctuations drive tissue cell irregularity, which increases $\langle s \rangle$ and $D_{eff}$. Second, though, fluctuations mitigate the effect of ECM tension, therefore \textit{reducing} the increase in $\langle s \rangle$ and $Q$ driven by external tension. These results are demonstrated in Fig. \ref{fig_supp_temp_vs_deff_shape_q}, (b) and (c), for fixed $p_{0}=3.9$. For $l_{0} > l_{0}^* \approx 0.62$, the ECM is floppy, and increasing the temperature results in an increase in $\langle s \rangle$ and little change in $Q$. However, for $l_{0} < l_{0}^*$, increasing $T$ attenuates the increase in $\langle s \rangle$ and decreases $Q$. In other words, at higher temperatures, higher ECM tensions are needed to observe the same $\langle s \rangle$ and $Q$. Despite the mitigating effect of temperature on $\langle s \rangle$ and $Q$, the effective diffusivity, $D_{eff}$, is not strongly effected by increasing temperature, as shown in Fig. \ref{fig_supp_temp_vs_deff_shape_q} (a), at least for the temperature range studied. 

\clearpage


%